\documentstyle[sprocl,epsf]{article}
\begin{document}

\bibliographystyle{unsrt}    

\def\Journal#1#2#3#4{{#1} {\bf #2}, #3 (#4)}

\def\NCA{\em Nuovo Cimento}
\def\NIM{\em Nucl. Instrum. Methods}
\def\NIMA{{\em Nucl. Instrum. Methods} A}
\def\NPB{{\em Nucl. Phys.} B}
\def\PLB{{\em Phys. Lett.}  B}
\def\PRL{\em Phys. Rev. Lett.}
\def\PRD{{\em Phys. Rev.} D}
\def\ZPC{{\em Z. Phys.} C}
\def\st{\scriptstyle}
\def\sst{\scriptscriptstyle}
\def\mco{\multicolumn}
\def\epp{\epsilon^{\prime}}
\def\vep{\varepsilon}
\def\ra{\rightarrow}
\def\ppg{\pi^+\pi^-\gamma}
\def\vp{{\bf p}}
\def\ko{K^0}
\def\kb{\bar{K^0}}
\def\al{\alpha}
\def\ab{\bar{\alpha}}
\def\be{\begin{equation}}
\def\ee{\end{equation}}
\def\bea{\begin{eqnarray}}
\def\eea{\end{eqnarray}}
\def\CPbar{\hbox{{\rm CP}\hskip-1.80em{/}}}

\def\mxfigura#1#2#3#4{
  \begin{figure}[hbtp]
    \begin{center}
      \epsfxsize=#1
      \leavevmode
      \epsffile{#2}
    \end{center}
    \caption{#3}
    \label{#4}
  \end{figure} }

\title{\bf CP-VIOLATION\footnote{Lectures delivered at the XXIV International
Meeting on Fundamental Physics, Playa de Gand\'{\i}a (1996)}}

\vspace{0.5cm}

\vspace{0.1cm}
\author{J. Bernab\'eu}

\vspace{0.2cm}

\address{
 Departamento de F\'{\i}sica Te\'orica, Univ. de Valencia\\
46100 Burjassot, Valencia, Spain}

\vspace{3cm}

\maketitle\abstracts{
 These lectures cover different aspects of the subject of
CP-Violation, from its description inside the Standard Model to the
phenomenological analysis of the $K^0 - \bar{K}^0$ system and the
prospects for its manifestation in B-physics.}

\begin{center}
{\bf Contents}
\end{center}

\vspace{0.2cm}

\noindent
1. CP in the Standard Model

1.1 Discrete symmetries: P,C 

1.2 Framework of local field theories

1.3 CPT Invariance: antiparticles

1.4 The Standard Electroweak Model

1.5 Yukawa Couplings

1.6 Spontaneous Symmetry Breaking

1.7 The Quark Mixing Matrix

1.8 CP-Violation with 3 families

\noindent
2. The $K^0 - \bar{K}^0$ System

2.1 Discovery of CP-Violation

2.2 Meson-Antimeson Mixing

2.3 Indirect CP-Violation

2.4 Isospin Decomposition

2.5 Experiments for $K_L \rightarrow 2 \pi$

2.6 CKM Quark Mixing Matrix

2.7 Coherent Decay of ($K^0, \bar{K}^0)$

2.8 Time Integrated Rates

\noindent
3. CP-Violation and B-Physics

3.1 Physics Motivation

3.2 Principle of the CP-violation measurement

3.3 Rate Asymmetries

3.4 Outlook

\newpage

\section{CP in the Standard Model}

\subsection{Discrete symmetries:P, C}

Conservation Laws in Physics are due to invariance of forces under 
symmetry transformations.

In particular, there are conservation laws corresponding to discrete
transformations: 

(i) Reflection in space or Parity (P). In three dimensions, it means
that the mirror image of an experiment yields the same results as
the original experiment. This  implies

\begin{center}
\begin{tabular}{|l|}
\hline\\
P-invariance means that ``left'' and ``right'' cannot be defined
in an \\
absolute sense.\\
\hline
\end{tabular}
\end{center}

\vspace{0.2cm}

(ii) Particle-antiparticle conjugation (C), i.e., C transforms each
particle into its antiparticle, without touching its space-time
properties. Similarly,

\begin{center}
\begin{tabular}{|l|}
\hline\\
C-invariance of laws means that experiments in a world of antimatter
will\\ 
give identical  results to the ones in our world.\\
\hline
\end{tabular}
\end{center}

\vspace{0.2cm}

As an example, C-invariance for electromagnetic interactions implies
that the atom of anti-hydrogen, recently discovered 
\cite{BAU} and manipulated
at CERN-LEAR, should show the same spectral lines as those
of the hydrogen atom.

Until the work of Lee and Yang \cite{LEE} in 1956 it was assumed that
all elementary processes are invariant under the separate symmetries $P$ and
$C$. Subsequent experiments in nuclear $\beta$-decay, $\pi^{\pm}$
decays and $\mu^\pm$ decays demonstrated the violation of $P$ and $C$
invariance under weak interactions. The application of these
symmetries to $\pi^\pm$ decays is shown in Fig.1, where the
configurations for momenta and helicities of the decay products are
exhibited for the $\pi^- \rightarrow \mu^- \bar{\nu}_\mu$
process and its $P, C$ and $CP$ transformed processes. Nature selects
that only the $\mu^-$helicity $+ 1$ in$\pi^-$-decay
and the $\mu^+$-helicity $- 1$ in $\pi^+$-decay are present, so $P$
and $C$ symmetries are violated whereas $CP$ is a good symmetry.

\newpage
\mxfigura{10cm}{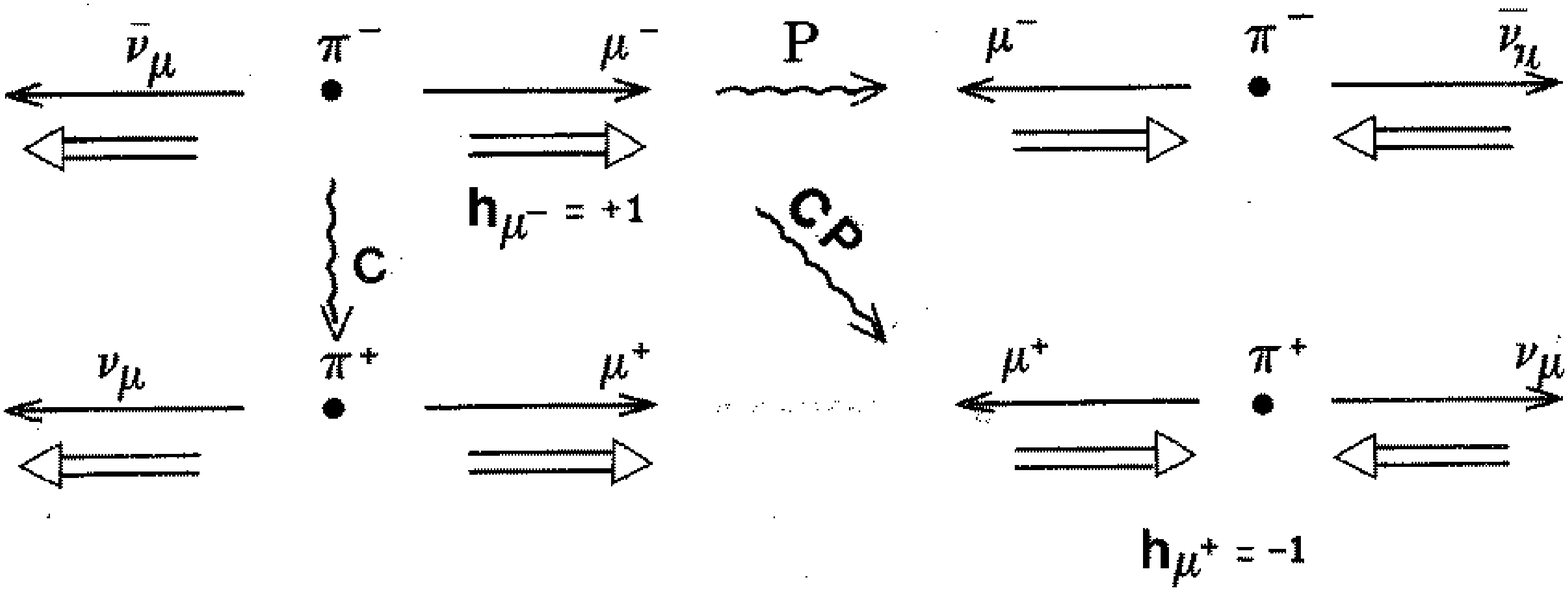}{C,P  and CP applied to $
\pi \rightarrow \mu \nu $}{figura1}

We conclude that

\begin{center}
\begin{tabular}{|l|}
\hline\\
The combined operation CP connects two physical (existing in \\
Nature!) processes.\\
\hline
\end{tabular}
\end{center}

\vspace{0.2cm}

CP-invariance was considered \cite{LAN} to be replacing the separate
$P$ and $C$ invariances of weak interaction. Not only for charge
current weak interactions: neutral current weak interactions, 
mediated by the $Z$ vector boson, are known at present to violate
$P$ and $C$ too.

P-Violation by neutral currents was beautifully demonstrated by
the left-right asymmetry in deep inelastic scattering of incoming 
electrons on a deuterium target at SLAC \cite{PRE}  and by
parity violation in atoms \cite{BOU}, where the interference of the
electromagnetic interaction with the $P$-odd neutral current interaction
operates. More recently, at the peak of the $Z$, the
left-right asymmetry for electron-positron collisions at SLC
\cite{SLD} measures the $P$-odd combination of neutral current couplings
of the electron, whereas the $P$-odd $\tau$ polarization \cite{LEP} has
been measured by LEP1. The forward-backward asymmetry in the
process $e^+ e^- \rightarrow f \bar{f}$ has been observed \cite{LEP}
first by LEP1 for both leptons and quarks in the final state
and it demonstrates C-violation in the neutral current interaction.

\subsection{Framework of local field theories}

In the framework of local field theories, one introduces a 
lagrangian density

$$
{\cal L} = {\cal L}_0 + {\cal L}_{int}
$$

\noindent
where ${\cal L}_0$ is the free part, describing noninteracting
particles, and ${\cal L}_{int}$ gives how these particles interact.
For $N$ fields $\phi_j (x)$, we assume

$$
{\cal L} (x) = {\cal L} (\phi_j (x), \partial^\mu \phi_j (x));
 \quad j = 1, 2, ..., N
$$

For the discussion of symmetries, the fundamental quantity is the
ACTION

$$
A = \int d^4 x {\cal L} (x) = A_0 + A_{int}
$$

\begin{center}
\begin{tabular}{|l|}
\hline\\
If the action $A$ is (is not) invariant under a symmetry operation,\\
then the symmetry is a good (broken) one.\\
\hline
\end{tabular}
\end{center}

\vspace{0.2cm}

We define the symmetry operation $P, C$ for the free fields 
entering ${\cal L}_0$ and ${\cal L}_{int}$: A field and
its P-or C-image satisfy the SAME EQUATION OF MOTION. One should
keep in mind that the fields in the Lagrangian density may or may
not correspond to physical fields (those describing particles
with well defined mass, lifetime, ...), a comment which is
particularly relevant in theories with spontaneous symmetry breaking.

\underline{Parity $P$} $(\vec{x} \rightarrow - \vec{x}) :
\vec{p} \rightarrow - \vec{p} , \vec{J} \rightarrow \vec{J}$

The free fields transform according to

$$
\begin{array}{ll}
& \hspace{1.6cm} P\\
\hbox{Scalar field} & \phi  (t,  \vec{x})   \rightarrow 
\phi (t, - \vec{x}) \\[2ex]
\hbox{Pseudoscalar field } & P (t,  \vec{x})\rightarrow 
- P (t, - \vec{x}) \\[2ex]
\hbox{Dirac spinor} & \psi (t,  \vec{x}) \rightarrow 
\gamma_0 \psi (t, - \vec{x}) \\
& \bar{\psi} (t,  \vec{x}) \rightarrow \bar{\psi} (t, - \vec{x}) 
\gamma_0\\[2ex]
\hbox{Vector field} & V_\mu (t,  \vec{x}) \rightarrow
V_\mu (t, - \vec{x})\\[2ex]
\hbox{Axial Field}& A_\mu (t,  \vec{x}) \rightarrow 
- A^\mu (t, - \vec{x})\\[2ex]
\end{array}
$$

Note that $\mu = 0, 1, 2, 3$ and, for any four vector $Q^\mu$,
one has $Q^0 = Q_0$, whereas $Q^k = - Q_k$ for $k = 1,2,3$.

The parity of a free field is not observable.
It is the interaction which fixes the ``relative'' parties of the
fields, if $P$ is a good symmetry.

For the Dirac spinors, the transformation properties of the bilinears
under $P$ are given by

$$
\begin{array}{ll}
& \hspace{1.3cm} P\\
& (t, \vec{x}) \rightarrow (t, - \vec{x})\\
\hbox{Scalar} & \bar{\psi}_1  \psi_2  \rightarrow 
\psi_1 \psi_2 \\[2ex]
\hbox{Pseudoscalar} & \bar{\psi}_1 \gamma_5 \psi_2 \rightarrow 
- \psi_1 \gamma_5 \psi_2 \\[2ex]
\hbox{Vector} & \bar{\psi}_1 \gamma_\mu \psi_2 \rightarrow 
\bar{\psi}_1 \gamma_\mu \psi_2 \\[2ex]
\hbox{Axial} & \bar{\psi_1} \gamma_\mu \gamma_5 \psi_2 \rightarrow
- \bar{\psi}_1 \gamma^\mu \gamma_5 \psi_2\\[2ex]
\hbox{Tensor}&  \bar{\psi}_1 \sigma_{\mu \nu} \psi_2 \rightarrow 
\bar{\psi}_1 \sigma^{\mu \nu} \psi_2\\[2ex]
\hbox{Pseudotensor} & \bar{\psi}_1
\sigma_{\mu \nu} \gamma_5 \psi_2 \rightarrow 
\bar{\psi}_1 \sigma^{\mu \nu} \psi_2\\ 
\end{array}
$$

These bilinears are very fundamental objects and appear frequently
in physics. From the transformation properties of the fields, written
in terms of the annnihilation and the creation  operators for
particles $a (\vec{p}, \lambda)$ and antiparticles $b (\vec{p}, \lambda)$,
one can find how the states $|\vec{p}, \lambda >$ transform.

\noindent
\underline{C-conjugation}: $a (\vec{p}, \lambda) \rightleftharpoons
b (\vec{p}, \lambda)$

The free fields have a Fourier decomposition in terms of the
annihilation operator $a (\vec{p}, \lambda)$ for particles plus the
creation operator $b^+ (\vec{p}, \lambda)$ for antiparticles. Under
C-conjugation, the role of the operators $a (\vec{p}, \lambda)$ and
$b (\vec{p}, \lambda)$ is interchanged. This exchange does not
necessarily mean that the \underline{physical} states of 
particles and antiparticles are connected by the C-operation, because the 
C-symmetry is violated in Nature. As an example, neutrinos and 
antineutrinos of the same helicity do not exist in Nature.

Under $C$, the free fields transform as follows

$$
\begin{array}{lll}
&\hspace{1cm} C\\
\hbox{Scalar field} & \phi (x)  \rightarrow \phi^+ (x)\\
\hbox{Dirac spinor} & \psi (x)  \rightarrow C \bar{\psi}^T (x)\\
& \bar{\psi} (x)  \rightarrow - \psi^T (x) C^{-1}\\
\hbox{Vector field} & V_\mu (x)  \rightarrow - V_\mu^+ (x)\\
\hbox{Axial field} & A_\mu (x)  \rightarrow A_\mu^+ (x)
\end{array}
$$

\noindent
where $C$ is a $4 \times 4$ unitary matrix which satisfies the condition

$$
C^{-1} \gamma_\mu C = - \gamma_\mu^T
$$

\noindent
as a result of imposing that the C-conjugated Dirac field satisfies the
same free field equation of motion as the field itself. The
matrix $C$ can be realized by the choice $C = i \gamma^2 \gamma^0$.

For a Real Vector Field, such as the photon $\gamma$ or the neutral
rho $\rho^0$, charged conjugation $C$ is well defined with eigenvalue
$C_\gamma = - 1$.

The spinor bilinears are transformed under $C$ as.

$$
\begin{array}{ll}
& x \quad \; \rightarrow \, x\\
\hbox{Scalar} & \bar{\psi}_1  \psi_2  \rightarrow 
\bar{\psi}_2 \psi_1 \\[2ex]
\hbox{Pseudoscalar} & \bar{\psi}_1 \gamma_5  \psi_2 \rightarrow 
 \bar{\psi}_2 \gamma_5 \psi_1 \\[2ex]
\hbox{Vector} & \bar{\psi}_1 \gamma_\mu \psi_2 \rightarrow 
- \bar{\psi}_2 \gamma_\mu \psi_1 \\[2ex]
\hbox{Axial} & \bar{\psi_1} \gamma_\mu \gamma_5 \psi_2 \rightarrow
 \bar{\psi}_2 \gamma_\mu \gamma_5 \psi_1\\[2ex]
\hbox{Tensor}&  \bar{\psi}_1 \sigma_{\mu \nu} \psi_2 \rightarrow 
- \psi_2 \sigma_{\mu \nu} \psi_1\\[2ex]
\hbox{Pseudotensor} & \bar{\psi}_1
\sigma_{\mu \nu} \gamma_5 \psi_2 \rightarrow 
- \bar{\psi}_2 \sigma_{\mu \nu} \gamma_5 \psi_1\\ 
\end{array}
$$

\noindent
\underline{CP-operation}: $(t, \vec{x}) \rightarrow (t, - \vec{x})$

$$
\begin{array}{ll}
\hbox{Scalar} & \bar{\psi}_1  \psi_2  \rightarrow 
\bar{\psi}_2 \psi_1 \\[2ex]
\hbox{Pseudoscalar} & \bar{\psi}_1 \gamma_5  \psi_2 \rightarrow 
 - \bar{\psi}_2 \gamma_5 \psi_1 \\[2ex]
\hbox{Vector} & \bar{\psi}_1 \gamma_\mu \psi_2 \rightarrow 
- \bar{\psi}_2 \gamma^\mu \psi_1 \\[2ex]
\hbox{Axial} & \bar{\psi}_1 \gamma_\mu \gamma_5 \psi_2 \rightarrow
 - \bar{\psi}_2 \gamma^\mu \gamma_5 \psi_1\\[2ex]
\hbox{Tensor}&  \bar{\psi}_1 \sigma_{\mu \nu} \psi_2 \rightarrow 
- \bar{\psi}_2 \sigma^{\mu \nu} \psi_1\\[2ex]
\hbox{Pseudotensor} & \bar{\psi}_1
\sigma_{\mu \nu} \gamma_5 \psi_2 \rightarrow 
 \bar{\psi}_2 \sigma^{\mu \nu} \gamma_5 \psi_1\\ 
\end{array}
$$

As an example, for an interaction Lagrangian of pseudoscalar-fermion-fermion
fields (think in $K_L \rightarrow \mu^+ \mu^-$) given by

$$
{\cal L}_{int} (x) = \bar{\psi} (x) [a + i b \gamma_5] \psi (x)
\phi (x)
$$

\noindent
the process violates parity $P$, conserves charge conjugation $C$ and
violates the combined $CP$. 

\subsection{ CPT Invariance: antiparticles}

A very important property of LOCAL FIELD THEORIES which respect
Lorentz invariance is that they are automatically invariant \cite{SCH}
under the combined operation CPT. This implies that the problem of the
invariance under CP is equivalent to that of the invariance under 
time-revesal $T$.

Taking into account the role of $T$ to transform the fourvectors
$x^\mu$ and $p^\mu$ and the angular momentum $\vec{J}$.

$$
(t, \vec{x}) \stackrel{T}{\longrightarrow} (- t, \vec{x})
; (p^0, \vec{p}) \stackrel{T}{\longrightarrow} (p^0,
- \vec{p}); \vec{J} \stackrel{T}{\longrightarrow} - \vec{J}
$$

\noindent
the spinor bilinears are transformed under CPT as follows

$$
\begin{array}{lccc}
 && CPT & \\
& (t, \vec{x}) & \rightarrow & (- t, - \vec{x})\\
& c-\hbox{number} & \rightarrow & (c-\hbox{number})^*\\[2ex]
\hbox{Scalar} & \bar{\psi}_1  \psi_2  & \rightarrow &
\bar{\psi}_2 \psi_1 \\[2ex]
\hbox{Pseudoscalar} & i \bar{\psi}_1 \gamma_5  \psi_2 & \rightarrow &
 i \bar{\psi}_2 \gamma_5 \psi_1 \\[2ex]
\hbox{Vector} & \bar{\psi}_1 \gamma_\mu \psi_2 & \rightarrow & 
- \bar{\psi}_2 \gamma_\mu \psi_1 \\[2ex]
\hbox{Axial} & \bar{\psi}_1 \gamma_\mu \gamma_5 \psi_2 & \rightarrow &
 - \bar{\psi}_2 \gamma_\mu \gamma_5 \psi_1\\[2ex]
\hbox{Tensor}&  \bar{\psi}_1 \sigma_{\mu \nu} \psi_2 & \rightarrow & 
 \bar{\psi}_2 \sigma_{\mu \nu} \psi_1\\[2ex]
\hbox{Pseudotensor} & i \bar{\psi}_1
\sigma_{\mu \nu} \gamma_5 \psi_2 & \rightarrow & i
 \bar{\psi}_2 \sigma_{\mu \nu} \gamma_5 \psi_1\\ 
\end{array}
$$

Since 1964 \cite{CHR} we know that, not only $C$ but, CP is also
a broken symmetry. This implies that the definition of the
physical states for antiparticles needs the use of the CPT-symmetry.

\subsection{The Standard Electroweak Model}

The present framework to discuss CP-violation is the Standard Model. The
Lagrangian density of the Electroweak Model is of the form

$$
{\cal L} = {\cal L} (f, G) + {\cal L} (f, H) + {\cal L} (G,H) +
{\cal L} (G) - V (H)
$$

\noindent
where $f =$ fermions (quarks, leptons), $G = $ gauge bosons
($\vec{W}$ and $B$),  $H = $the scalar doublet.

The Lagrangian is constructed so that it is invariant under the local
(space-time dependent) symmetry group $SU (2) \times U (1)$. Under
$SU (2)$, the quark fields transform as doublets, if they are
left-handed, and as singlets if they are right-handed. One introduces
the multiplets

$$
\left\{ \begin{array}{c}
q_j\\
q'_j \end{array} \right\}_L \quad ; \quad q_{jR} \; \,, \; \,q'_{jR} \quad ;
\quad \quad j = 1, 2, ... N
$$

\noindent
where the index $j$ is the family index and $N$ denotes the number
of families.

The hadronic part of the Lagrangian between fermions and gauge
bosons is

$$
\begin{array}{c}
{\cal L} (f, G) = \sum_{j = 1}^N \left\{(\overline{q, q'})_{jL}
i \gamma^\mu \, [\partial_\mu - i g_2 \frac{\vec{\sigma}}{2} \cdot
\vec{W}_\mu - i g_1 (\frac{1}{6}) B_\mu] \right. 
\{ \begin{array}{c}
q\\
q'\end{array} \}_{jL} \\[2ex]
+ \bar{q}_{jR} i \gamma^\mu \, [\partial_\mu - ig_1 (\frac{2}{3}) B_\mu]
q_{jR} + \bar{q}'_{jR} i \gamma^\mu \, [\partial_\mu - i g_1 (-
\frac{1}{3}) B_\mu] \, q'_{jR} \left. \right\}
\end{array}
$$

\noindent
where the numbers in the parenthesis are the eigenvalues of the weak
hypercharge $Y$, the generator of the $U (1)$ group. They are chosen
such that $Q \equiv I_3 + Y$ has the eigenvalues $\frac{2}{3}$ for the
up-type quarks and $- \frac{1}{3}$for the down-type quarks.

The Lagrangian density ${\cal L} (f,G)$ violates both $P$ and $C$
symmetries. For $P$, we observe that the interaction of left-handed
quarks is different from the interaction of right-handed quarks.
This $P$-non invariance remains even after the lagrangian is rewritten
in terms of the physical fields. The simultaneous presence of vector
\underline{and} axial interactions leads to the violation of C-symmetry, 
because the vector density is odd, whereas the axial density is even.
However, ${\cal L} (f, G)$ is CP-invariant \cite{GRI}.

\subsection{Yukawa Couplings}

The Lagrangian density ${\cal L} (G, H)$ between the gauge bosons
and the scalars is both P- and C-symmetric.

\begin{tabular}{|l|}
\hline\\
All CP-violation, in the Electroweak Standard Model, originates from\\
${\cal L} (f, H)$.\\
\hline
\end{tabular}
\vspace{0.2cm}

The hadronic part of the Yukawa interaction between fermions and 
scalars is given by

$$
\begin{array}{cc}
{\cal L} (f, H) = & \sum_{j, k = 1}^N \left\{ Y_{jk} (\overline{q, q'})_{jL}
\right. \left\{
\begin{array}{c}
\phi^{(0)*}\\
- \phi^{(-)} \end{array} \right\} q_{kR} \\[2ex]
 & + \quad Y'_{jk} (\overline{q, q'})_{jL}
\left\{
\begin{array}{c}
\phi^{(+)}\\
\phi^{(0)} \end{array} \right\} q'_{kR} 
 \quad + \,  \hbox{h.c.} \left. \right\}
\end{array}
$$

The Yukawa couplings $Y_{jk}, Y'_{jk}$ are arbitrary complex
numbers, because  ${\cal L} (f, H)$ is manifestly invariant under
the gauge group $SU (2) \times U (1)$ symmetry. The Lagrangian density
${\cal L} (f, H)$ involves scalar and pseudoscalar interactions, so
it violates  $P, C$ and $CP$ symmetries. However, in this case it is of
relevance to discuss the symmetry properties after spontaneous 
symmetry breakdown.

\subsection{Spontaneous Symmetry Breaking}

Under the spontaneous symmetry breaking, in the unitary gauge, the
complex field $\phi^{(0)}$ is shifted and becomes real and
the field $\phi^{(+)}$ vanishes:

$$
\left.
\begin{array}{cc}
\phi^{(0)} \rightarrow & \frac{1}{\sqrt{2}} (v + H)\\
\phi^{(+)} \rightarrow & 0 \end{array} \right\}
\begin{array}{ll}
v \equiv & \hbox{v.e.v.}\\
H \equiv & \hbox{Higgs field}
\end{array}
$$

\vspace{0.2cm}

\begin{tabular}{|l|}
\hline\\
From the four degrees of freedom of the scalar doublet, 
three fields
are \\"eaten" by the longitudinal components of the
$W^\pm$ and $Z$ bosons,
which \\ become massive.\\
\hline
\end{tabular}

\vspace{0.2cm}
 
After spontaneous symmetry breaking, what remains for
${\cal L} (f, H)$ is 

$$
{\cal L} (f, H) \stackrel{SSB}{\longrightarrow} - \sum_{j, k=1}^{N}
\left\{ m_{jk} \bar{q}_{jL} q_{kR} + m'_{jk}
\bar{q}'_{jL} q'_{kR} \quad  + \, \hbox{h.c.} \right\}
\quad [1 + \frac{1}{v} H]
$$

\noindent
where the quantities

$$
m_{jk} = - \frac{v}{\sqrt{2}} Y_{jk} \quad , \quad
m'_{jk} = - \frac{v}{\sqrt{2}} Y'_{jk}
$$

\noindent
are the complex QUARK MASS MATRICES. Their dimension is $N \times N, N$
being the number of families. This Lagrangian violates the discrete
symmetries $P$, $C$ and $CP$:

\vspace{0.2cm}
\begin{tabular}{llll}
$P$ & is violated if & $[m], [m']$ & are not hermitian.\\
$C$ & is violated if & $[m] , [m']$  & are not symmetric.\\
$CP$ &  is violated if & $[m], [m']$  & are not real.\\
\end{tabular}

\vspace{0.2cm}

However, we should keep in mind that we are, so far, dealing with 
non-physical fields. In order to find the physical fields we must 
diagonalize the quark mass matrices $[m], [m']$. We know that
any square matrix, hermitian or not, can be diagonalied by means
of two unitary matrices:

$$
\begin{array}{lllll}
U_L &  m U_R^+ &  = &  D & \equiv \hbox{Diag} (m_u, m_c, m_t)\\
U_L' &   m' U'^+_R &  = &  D' & \equiv \hbox{Diag} (m_d, m_s, m_b)
\end{array}
$$

We can discover the meaning of $U_L$ as

$$
U_L \, m \, U_R^+ \, U_R \, m^+  \, U_L^+ = U_L \, m \, m^+ \, U_L^+ = D^2
$$

\noindent
the unitary matrix which diagonalizes the hermitian matrix $m m^+$.
Similarly, $U_R$ is the unitary matrix diagonalizing $m^+ m$. Analogous
relations can be derived for the down-quark sector, putting primes
on the $U' s$.

In order to identifly the physical fields, we write from the Lagrangian
density

$$
\begin{array}{l}
\bar{q}_{jL} m_{jk} q_{kR} \equiv \bar{q}_L m q_R = \bar{q}_L U^+_L
U_L m U^+_R U_R q_R\\[2ex]
\quad = \overline{U_L q_L} D U_R q_R = \overline{U_L q_L}
\left( \begin{array}{ccc}
m_u & 0 & 0\\
0 & m_c & 0\\
0 & 0 & m_t \end{array} \right)  U_R q_R
\end{array}
$$

Thus the physical fields are

$$
q_L^{phys} = U_L q_L = U_L \left\{ \begin{array}{l}
u_L\\ c_L \\ t_L \end{array} \right\} \quad ; \quad q'^{phys}_{L}
= U'_L q'_L = U'_L  \left\{ \begin{array}{l}
d_L\\
s_L\\ 
b_L 
\end{array} \right\}
$$

\noindent
with similar relations valid for the R-handed quarks. The Yukawa
Lagrangian becomes

$$
\begin{array}{ll}
{\cal L}^{phys} (f, H) & = - (1 + \frac{H}{v}) 
[m_u \bar{u}  \frac{1 + \gamma_5}{2} u + m_c  \bar{c} \frac{1 + \gamma_5}{2}
c + ...\\[2ex]
& + m_d \bar{d} \frac{1 + \gamma_5}{2} 
d + m_s \bar{s} \frac{1 + \gamma_5}{2} s + ... ] \quad + \hbox{h.c.}\\[2ex]
& = - (1 + \frac{H}{v}) [ m_u \bar{u} u + m_c
\bar{c} c + ...\\[2ex]
& \hspace{2.1cm} + m_d \bar{d} d + m_s \bar{s} s + ... ]
\end{array}
$$

The essential feature of ${\cal L}^{phys} (f, H)$ is that it conserves,
separately, $P$ and $C$, and thus also $CP$ and $T$. The lesson
to be learnt \cite{JAR} from this exercise is that the "apparent"
symmetry properties of a Lagrangian density need not have anything to
do with Physics, when this Lagrangian is expressed in terms of
unphysical fields.

At this point, it is of interest to summarize our findings for the Standard
Model:

\vspace{0.2cm}
\begin{tabular}{|l|}
\hline\\
1) Given the quark mass matrices $m, m'$, we need four unitary matrices\\
$U_L , U_R \; ; \; U'_L , U'_R$\\
to diagonalize them.\\
2) The unitary diagonalizing matrices allow to find the relations
between\\ 
physical (eigenfields of mass) and nonphysical (fields
 with definite \\
transformation  properties under the gauge group) fields.\\
3) The Lagrangian density ${\cal L}^{phys} (f, H)$, in terms
of physical fields, has no \\
traces of the unitary $U, U'$ matrices.\\
4) The physical Higgs-fermion interaction conserves $P, C$ and it
is Flavour \\
Diagonal.\\
5) The above properties 1) $\rightarrow $ 4) could change if one
introduces Physics \\Beyond the Standard Model.\\
\hline
\end{tabular}

\vspace{0.2cm}

\subsection{The Quark Mixing Matrix}

Now that we have the term ${\cal L} (f, H)$, in terms of the physical
quark fields, we turn our attention to the other term of the Lagrangian
density which needs to be expressed in terms of the physical
quark fields, i.e. ${\cal L} (f, G)$. Consider first any "neutral
current" term, i.e., any term which involves only either the 
up-kind quarks or the down-kind quarks but not both. As an  example.

$$
\begin{array}{ll}
\bar{q}_{jR} \gamma^\mu & [\partial_\mu - i g_1 (\frac{2}{3})
B_\mu ] q_{jR} = \\[2ex]
& = \bar{q}^{phys}_R U_R \gamma^\mu [\partial_\mu - i g_1 (\frac{2}{3})
B_\mu ] U_R^+ q_R^{phys}\\[2ex]
& = \quad \bar{q}^{phys}_R \gamma^\mu [\partial_\mu - i g_1 (\frac{2}{3})
B_\mu ] q_R^{phys}
\end{array}
$$

As a consequence of unitarity $U_R U^+_R = 1$ and $U_R$ again
"disappears": neutral Current terms are Flavour Conserving. In the
Standard Model there are some "miracles" which happen 
"naturally" [see  \cite{GLA} ]:

i) Neutral Currents all conserve CP.

ii) There are No Flavour-Changing-Neutral-Currents, due to two facts

\hspace{2cm} - these terms are helicity conserving

\hspace{2cm} - the four matrices $U, U'$ (with indices $L$ or $R$) are
unitary.

 This absence of flavour changing neutral currents  is usually 
referred to as the GIM mechanism \cite{LGL}. As there are stringent limits,
from experiment, on FCNC, the absence of such currents in the Standard
Model is one of its great successes. One can note that Physics
Beyond the Standard Model generally encounters difficulties in this
respect.

What remains to be treated in ${\cal L} (f,G)$ is the charged current
term. Such terms are helicity conserving, but \underline{do mix}
the up-type and the down-type quarks, simply because $W^\pm$ carry one unit 
of charge. Since the charged currents only involve left-handed quarks 
the matrices $U_R, U'_R$ do not enter, but $U_L \; U_L'^{ +} \neq
I$.

The charged current terms are given by

$$
\begin{array}{l}
[W^1_\mu -i W^2_\mu] \bar{q}_L \gamma^\mu q'_L + \, \hbox{h.c.} \, =
[W_\mu^1 -i W_\mu^2] \bar{q}^{phy}_L \gamma^\mu U_L U'^+_L
q'^{phys}_L \quad  + \hbox{h.c.} = \\[2ex]
[W^1_\mu -i W^2_\mu] \bar{q}_L^{phys} \gamma^\mu V q'^{phys}_L
 + \quad   \hbox{h.c.} \, \equiv
[W_\mu^1 -i W_\mu^2] J^\mu_{cc}  \quad + \hbox{h.c.}
\end{array}
$$

\noindent
where $V$ is a unitary matrix and

$$
J^\mu_{cc} \equiv [\bar{u}, \bar{c}, \bar{t}]_L \gamma^\mu
\left(
\begin{array}{lll}
V_{ud} & V_{us} & V_{ub}\\
V_{cd} & V_{cs} & V_{cb}\\
V_{td} & V_{ts} & V_{tb}\\ \end{array} \right)
\left\{ \begin{array}{l}
d\\
s\\
b \end{array} \right\}_L
$$

The matrix $V$ is the Quark Mixing Matrix. 

We note the following properties:

\begin{center}
\begin{tabular}{|l|}
\hline\\
1) All CP-Violation is in $V$, together with Flavour Mixing.\\
2) Charged Currents violate $P,C$ maximally.\\
3) CP-symmetry would require $V$ real, modulus unmeasurable phases.\\
\hline
\end{tabular}
\end{center}

\subsection{CP-Violation with 3 Families}

What matters in field theory is not the absolute phases but the relative
phases of different fields. Which phases in $V$ are measurable ? Under
rephasing of the L-quark fields, one finds

$$
V_{\alpha j} \rightarrow e^{i [\phi (j) - \phi (\alpha)]} V_{\alpha j}
$$

\noindent
where $\alpha (j)$ denotes and up-kind (down-kind) quark.

This rephasing of L-fields only affects the term ${\cal L}^{phys} (f, H)$
in the Lagrangian density, but it can be remedied by rephasing the 
R-fields, so that the Lagrangian density remains invariant. A general
unitary $V$ matrix has $N^2$ parameters, with

$$
\frac{N (N- 1)}{2} \, \hbox{moduli} \quad , \quad
\frac{N (N + 1)}{2}  \, \hbox{phases}
$$

From rephasing invariance, we can absorb $(2N - 1)$ unobservable phases,
so that \cite{BER} one finishes with $(N - 1)^2$ physical parameters, among
which

$$
\frac{N (N- 1)}{2} \, \hbox{moduli} \quad , \quad
\frac{(N - 1) (N - 2)}{2}  \, \hbox{phases}
$$

For the case of 2 families, $N = 2$, we find 1 rotation angle and
no phases. The quark mixing matrix is then the Cabibbo matrix \cite{CAB},
which is real orthogonal. Reality implies that $CP$ is
conserved in the Standard Model with 2 families.

For 3 families, $N = 3$, the number of physical parameters in $V$ is 4,
i.e., 3 rotation angles and 1 phase. The first example of such a 
matrix was given by Kobayashi and Maskawa \cite{KOB}.

In order to have CP-violation, the three families have to be active
in the process at hand. If two quarks with the same charge were degenerate
in mass, there is an extra symmetry

$$
(s, b) \rightarrow s^{new} = V_{us} s + V_{ub} b
$$

\noindent
such that $u$ is not coupled to $b^{new}$, and the new $V$ could
be chosen real. Necessary conditions for CP-violation are

$$
\begin{array}{l}
m_u \neq m_c \quad , \quad m_c \neq m_t \quad , \quad m_t \neq
m_u\\
m_d \neq m_s \quad , \quad m_s \neq m_b \quad , \quad m_b \neq
m_d\\
\end{array}
$$

There are unfortunately many possible parametrizations of $V$.
PDG advocates \cite{PAR} the following form

$$
V = \left[
\begin{array}{ccc}
c_{12} c_{13} & s_{12} c_{13} & s_{13} e^{-i \delta_{13}}\\
-s_{12} c_{23} - c_{12} s_{23} s_{13} e^{i \delta_{13}} &
c_{12} c_{23} -s_{12} s_{23} s_{13} e^{i \delta_{13}} & s_{23} c_{13}\\
s_{12} s_{23} -c_{12} c_{23} s_{13} e^{i \delta_{13}} &
- c_{12} s_{23} - s_{12} c_{23} s_{13} e^{i \delta_{13}} &
c_{23} c_{13}
\end{array}
\right]
$$

\noindent
where the rotation angles $\theta_{12}, \theta_{23}, \theta_{13}$ can be 
made to lie in the first quadrant, so that

$$
c_{ij} \geq  0 \quad , \quad s_{ij} \geq 0 \quad ; \quad 0 \leq \delta_{13}
\leq 2 \pi
$$

A very interesting approximate parametrization is due to Wolfenstein 
\cite{WOL},
in terms of the four parameters

$$
(\lambda, A, \rho, \eta)
$$

\noindent
with $\lambda$, the Cabibbo angle, taken as the expansion parameter:

$$
V = \left[
\begin{array}{ccc}
1 -  \frac{1}{2} \lambda^2 & \lambda & \lambda^3 A (\rho - i \eta)\\
- \lambda &  1 - \frac{1}{2} \lambda^2 & \lambda^2 A\\
\lambda^3 A (1 - \rho - i \eta) & - \lambda^2 A & 1
\end{array}
\right]
$$

In this parametrization, CP-violation is manifested in the presence
of an imaginary part of the $V_{ub}, V_{td}$ matrix elements: $\eta \neq 0$.

\section{The $K^0 - \bar{K}^0$ system}

\subsection{Discovery of CP-Violation}

One consequence of CP-INVARIANCE for the neutral $K^0 - \bar{K}^0$ system
was predicted by Gell-Mann and Pais \cite{GEL}: there should be
a long-lived partner to the known $V^0 (K_1^0)$ particle of short
lifetime: $\tau (K_s)\sim 10^{-10}$sec. According to this proposal these
two particles are coherent superpositions of two strangeness
eigenstates, $K^0 (S = +1)$ and $\bar{K}^0 (S = -1)$, produced in
interactions with conservation of strangeness, like strong interactions.
Weak interactions do not conserve strangeness and the physical
particles should be eigenstates of CP if the weak interactions are
CP-invariant. These eigenstates are (with $\bar{K}^0 \equiv - CP \; K^0$):

$$
\begin{array}{l}
CP \, K_1 = CP \frac{1}{\sqrt{2}} (K^0 - \bar{K}^0) = \frac{1}{\sqrt{2}}
(- \bar{K}^0 + K^0) = K_1\\[2ex]
CP \, K_2 = CP \frac{1}{\sqrt{2}} (K^0 + \bar{K}^0) = \frac{1}{\sqrt{2}}
(- \bar{K}^0 - K^0) = - K_2
\end{array}
$$

\noindent
$K_1, K_2$, eigenstates of CP, should be the physical particles with 
definite mass and lifetime.

Because of $CP (\pi^+ \pi^-) = (\pi^+ \pi^-), CP
(\pi^0 \pi^0) = ( \pi^0 \pi^0)$, the decay into ($\pi \pi)$ is
allowed for $K_1$, to be identified with the experimental state $K_s$
of short lifetime. The decay into $(\pi \pi)$ is forbidden for $K_2$,
hence its longer lifetime, $K_L$, which was indeed confirmed when
discovered \cite{BAR}.

\begin{center}
\begin{tabular}{|l|}
\hline\\
In 1964, Christenson, Cronin, Fitch and Turlay \cite{CHR} discovered
that 
$K_L$ \\
also decays to $\pi^+ \pi^-$ with a branching ratio of
$\sim 2 \times 10^{-3}$. 
This discovery \\established CP-VIOLATION.\\
\hline
\end{tabular}
\end{center}

\vspace{0.2cm}

Then, the long-lived neutral Kaon $K_L$ is no longer identical to the
CP-eigenstate $K_2$. Similarly, $K_s$ is not identical to $K_1$. 
CP-Violation,
discovered by the decay $K_L \rightarrow \pi^+ \pi^-$, was confirmed
later by
$K_L \rightarrow \pi^0 \pi^0 $ \cite{GAI} and by the charge asymmetry
\cite{BEN} in the $K_{l3}$ decays: $K_L \rightarrow
\pi^\pm e^\mp \nu$ and $K_L \rightarrow \pi^\pm \mu^\mp \nu$.

Early models for CP-violation atributed this phenomenon to a fifth force,
the superweak interaction \cite{WOF}, to a T-odd part of the weak
interaction or to the interference with a T-odd part of the 
electromagnetic or strong interaction. Many subsequent experiments
have reduced the possible models to the superweak and milliweak classes.
Amongst the milliweak models, the Present Orthodoxy is based on
the Complex Quark Mixing Matrix for 3 families

\subsection{Meson-Antimeson Mixing}

Let the eigenstates of flavour be $P^0 (= K^0, D^0, B^0)$
and $\bar{P}^0$. Flavour Number is not conserved by $H_w$ of weak
interactions, so in 2$^{nd}$ order $P^0$ and $\bar{P}^0$ can
mix through the intermediate states connected by $H_w$. The mixing
leads to

$$
| \psi (t) > = a (t) | P^0 > + b (t)
| \bar{P}_0 > \equiv \left\{ 
\begin{array}{l}
a (t)\\
b (t) \end{array}
 \right\}
$$

\noindent
with the Time Evolution governed by the Mixing Matrix ${\cal M}$. Assuming
CPT-invariance,

$$
\begin{array}{c}
i \frac{d}{dt} | \psi (t) > = {\cal M} | \psi (t) > \\[2ex]
{\cal M} = \left( \begin{array}{cc}
M & M_{12}\\
M_{12}^* & M \end{array} \right)
- \frac{i}{2} \left( \begin{array}{cc}
\Gamma  & \Gamma_{12}\\
\Gamma_{12}^* & \Gamma
\end{array} \right)
\end{array}
$$

The non-hermiticity of ${\cal M}$  is associated with the existence
of decay channels, not introduced explicitely in the formalism. The
values of $M$ and $\Gamma$, real, would correspond to the mass and 
width of both $P^0, \bar{P}^0$ in absence of Mixing.

The off-diagonal entries $M_{12}, \Gamma_{12}$ of the mass matrix 
${\cal M}$ correspond to the Dispersive and Absorptive parts, 
respectively, of the $\Delta P = 2$ transitions. If CP were a good
symmetry, then $M_{12}, \Gamma_{12}$ would be real.

The physical eigenstates of ${\cal M}$ are

$$
| P_\mp > = \frac{1}{\sqrt{|p|^2 + |q|^2}} \, [ p | P^0 > \mp \, 
q | \bar{P}^0 >]
$$

\noindent
with the amplitudes

$$
\frac{q}{p} \equiv \frac{1 - \bar{\varepsilon}}{1 + \bar{\varepsilon}} =
\left( \frac{M_{12}^* - \frac{i}{2} \Gamma_{12}^*}{M_{12} - \frac{i}{2}
\Gamma_{12}} \right)^{1/2}
$$

In the limit of CP-symmetry, $M_{12}$ and $\Gamma_{12}$ are real
and thus $q/p = 1$: the physical eigenstates become eigenstates of $CP$

$$
| P_{1,2} > = \frac{1}{\sqrt{2}} (|P^0 > \mp | \bar{P}^0 >) \; ; \; 
CP |P_{1,2} >  = \pm | P_{1,2} >
$$

In the presence of CP-violation, the physical eigenstates $|P_\pm >$
are NOT orthogonal, a result implied by the non-hermiticity of ${\cal M}$.
One has 

$$
< P_- | P_+ > = \frac{|p|^2 - |q|^2}{|p|^2 + |q|^2} \approx 2 \; Re
\; \bar{\varepsilon}
$$

With the initial condition ($t = 0)$ that the produced state is either
$P^0$ or $\bar{P}^0$, the corresponding Time Evolution gives

$$
\left\{ \begin{array}{l}
|P^0 (t) >\\
| \bar{P}^0 (t) > \end{array} \right\} = 
\left[ \begin{array}{cc}
g_1 (t) & \frac{q}{p} g_2 (t)\\
\frac{p}{q} g_2 (t) & g_1 (t) \end{array} \right]
\left\{ \begin{array}{l}
| P^0 > \\
| \bar{P}^0 > \end{array} \right\}
$$

\noindent
where the two time-dependent functions $ g_{1,2} (t)$
are

$$
\left\{ \begin{array}{c}
g_1 (t)\\
g_2 (t) \end{array} \right\} = e^{-i Mt} e^{- \Gamma t/2}
\left\{ \begin{array}{l}
\cos (\Delta M - \frac{i}{2} \Delta \Gamma ) \frac{t}{2} \\
- i \sin (\Delta M - \frac{i}{2} \Delta \Gamma) \frac{t}{2}
\end{array} \right\}
$$

\noindent
with $\Delta M \equiv M_{P_+} - M_{P_-} \quad ; \quad 
\Delta \Gamma \equiv \Gamma_{P_+} - \Gamma_{P_-}$

The main difference between the $K^0 - \bar{K}^0$ system and the
$B^0 - \bar{B}^0$ system lies in the relative values
of these parameters:

$$
\begin{array}{lll}
- K \; \hbox{system}, & \Gamma_{K_L} < < \Gamma_{K_s}, & \Delta
\Gamma_{K} \approx - \Gamma_{K_s} \approx - 2 \Delta M_K\\
- B \; \hbox{system}, & \Delta \Gamma_B < < \Gamma_B, & \Delta \Gamma_B
<< \Delta M_B
\end{array}
$$

\subsection{Indirect CP-Violation}

The flavour-specific decays 

$$
\begin{array}{l}
K^0 \rightarrow \pi^- l^+ \nu_l\\
\bar{K}^0 \rightarrow \pi^+ l^- \bar{\nu}_l
\end{array}
$$

\noindent
can measure whether $| p/q | \neq 1$, i.e., CP-violation in the physical
eigenstates of the Mass Matrix, referred to as Indirect CP-violation.

In the Standard Model, the semileptonic decay amplitudes from $K^0$
or $\bar{K}^0$ are equal

$$
| A (\bar{K}^0 \rightarrow \pi^+ l^- \bar{\nu}_l)| = | A (K^0 \rightarrow
\pi^- l^+ \nu_l)|
$$

\noindent
so that the charge asymmetry from $K_L$-decays is given by

$$
\delta = \frac{ \Gamma (K_L \rightarrow \pi^- l^+ \nu_l) - 
\Gamma (K_L \rightarrow \pi^+ l^- \bar{\nu}_l)}{
\Gamma (K_L \rightarrow \pi^- l^+ \nu_l) + 
\Gamma (K_L \rightarrow \pi^+ l^- \bar{\nu}_l)} = 
\frac{|p|^2 - |q|^2}{|p|^2 + |q|^2}
$$

\noindent
and indicates that all the CP-violating effect comes from the selection of
either $K^0$ or $\bar{K}^0$ in the
$K_L$-state.

Present values of $\delta$ give a world average \cite{PAR}

$$
\delta = (3.27 \, \pm \, 0.12) \times 10^{-3}
$$

\noindent
so that there is

\begin{center}
\begin{tabular}{|l|}
\hline\\
CP-Violation in the Mixing \\
$Re \, \bar{\varepsilon}_K
= (1.63 \pm 0.06) \times 10^{-3}$\\
\hline
\end{tabular}
\end{center}

\subsection{Isospin Decomposition}

For the non-leptonic $K_{S,L} \rightarrow 2 \pi$ decays, the
angular momentum of the pions  vanishes. The spatial part of the 
$(2 \pi)$-state is therefore symmetric, and since pions are bosons, the 
isospin state must be symmetric too. The two symmetric combinations
of two $I = 1$ states have $I = 0$ and $I = 2$, and the four 
existing transition amplitudes are 

$$
< 0 |T| K_s > \quad , \quad < 2 |T| K_s > \quad , \quad < 0 |T| K_L >
\quad , \quad < 2 |T| K_L >
$$

By normalizing to the dominant CP-conserving $\Delta I = 1/2$ transition
amplitude $< 0 |T| K_s>$, we define three complex numbers

$$
\begin{array}{l}
\varepsilon_0 \equiv < 0 |T| K_L > / < 0 |T| K_s >\\
\varepsilon_2 \equiv  \frac{1}{\sqrt{2}} < 2 |T| K_L > / < 0 |T|
K_s >\\
\omega \equiv < 2 |T| K_s > / < 0 |T| K_s >
\end{array}
$$

Non-vanishing values of $\varepsilon_{0,2}$ are a demonstration of 
CP-violation, whereas $\omega$ parametrizes the relative  contribution
of the $\Delta I = 3/2$ amplitude.

The experimentally observable quantities are

$$
\begin{array}{l}
\eta_{+ -} = < \pi^+ \pi^- |T| K_L > / < \pi^+ \pi^- |T| K_s>\\
\eta_{0 0 } \hspace{0.2cm} = < \pi^0 \pi^0 |T| K_L > / < 
\pi^0 \pi^0 |T| K_s>\\
\end{array}
$$

Relating the isospin states to the physical ($2\pi$)-states,

$$
\begin{array}{l}
< 0 | = \frac{1}{\sqrt{3}} < \pi^- \pi^+ | - \frac{1}{\sqrt{3}}
< \pi^0 \pi^0 | + \frac{1}{\sqrt{3}} < \pi^+ \pi^- |\\
< 2 | = \frac{1}{\sqrt{6}} < \pi^- \pi^+ | + \sqrt{\frac{2}{3}}
< \pi^0 \pi^0 | + \frac{1}{\sqrt{6}} < \pi^+ \pi^- |
\end{array}
$$

\noindent
one obtains

\[ {
\fbox{
$
\eta_{+ -} = \frac{\displaystyle{
\varepsilon_0 + \varepsilon_2}}{\displaystyle{1 + \frac{\omega}{\sqrt{2}}}}
\quad , \quad \eta_{00} = \frac{\displaystyle{
\varepsilon_0 - 2 \varepsilon_2}}{\displaystyle{1 -
\sqrt{2} \omega }}
$ }}
\]

Because of the validity of the $\Delta I = 1/2$ rule for 
CP-conserving weak non-leptonic decays, one expects, $|\omega| << 1$.
Its actual value gives $| \omega | \sim 1/22$.

A suitable choice for the phase of the $K^0 \rightarrow 2 \pi \;
(I = 0)$ amplitude is obtained by putting all its complexity as coming
from final-state-interactions leading to a phase-shift $\delta_0$:

$$
\left.
\begin{array}{l}
< 0 |T| K^0 > \equiv i A_0 e^{i \delta_0}\\
 < 0 |T| \bar{K}^0 > \equiv -i A^*_0 e^{i \delta_0}
\end{array} \right\}
\begin{array}{c}
A_o real,\\ 
Im (A_0) = 0 \end{array}
$$

Similarly, one has 

$$
\left.
\begin{array}{l}
< 2 |T| K^0 > \equiv i A_2 e^{i \delta_2}\\
< 2 |T| \bar{K}^0 > \equiv -i A_2^* e^{i \delta_2}
\end{array}
\right\}
$$

Then one gets

$$
\varepsilon_0 = \varepsilon_K \quad , \quad \varepsilon_2 = i
\frac{1}{\sqrt{2}} \frac{Im A_2}{A_0} e^{i (\delta_2 - \delta_0)}
\equiv \varepsilon '
$$

\noindent
as well as the charge asymmetry $\delta = 2 \, Re \, \varepsilon_K$

\vspace{0.2cm}

\begin{tabular}{|l|}
\hline\\
We observe:\\
(i) $\varepsilon_K$ measures indirect CP-violation\\
(ii) $\varepsilon '$ measures direct CP-violation, in the decay
amplitude, governed\\
by the phase-shift in final-state-interaction \cite{PAR}
$\delta_2 - \delta_0 = - 45^{o} \pm 6^o$\\
(iii) The observable quantities satisfy the Wu-Yang triangle relations\\
$\eta_{+ -} = \varepsilon + \varepsilon' \quad , \quad
\eta_{00} = \varepsilon - 2 \varepsilon '$\\
The experimental objective is the separation of $\varepsilon$ and
$\varepsilon'$\\
\hline
\end{tabular}

\vspace{0.2cm}

The phase of $\varepsilon '$ is $Arg (\varepsilon ') = \frac{\pi}{2}
+ (\delta_2 - \delta_0) = (45^0 \pm 6^0$).

The phase of $\varepsilon$ can be identified from the diagonalization
of the mass matrix ${\cal M}$.

$$
\varepsilon_K \simeq e^{i \phi_{sw}} \frac{Im M_{12} - \frac{i}{2}
Im \Gamma_{12}}{\sqrt{\Delta M_K^2 + \frac{1}{4} \Delta \Gamma^2_K}}
$$

\noindent
where  $\phi_{sw}$ is the so-called superweak phase

$$
\phi_{sw} \equiv  \arg \tan (\frac{- 2 \Delta M_K}{\Delta
\Gamma_K})
$$

Taking into account $\Delta \Gamma_K \simeq - 2 \Delta M_K$, we discover
that $\phi_{sw}$ is again near to $\pi /4$. More precisely,
$\phi_{sw} = 43.64^0 \pm 0.15^0$.

The phase introduced by the second factor in $\varepsilon_K$ is very
small, so one concludes that the phases of  $\varepsilon$ 
and $\varepsilon '$ are nearly
equal. The equality of the phases $\phi_{+ -}$ and $\phi_{00}$ of the
observable amplitudes $\eta_{+ - }$ and $\eta_{00}$, respectively, is 
a test of CPT-invariance in Nature.

\subsection{ Experiments for $K_L \rightarrow 2 \pi$}

The experimental results

$$
| \eta_{+-} | = (2.269 \pm 0.023) \times 10^{-3} \, , \, 
| \eta_{00} | = (2.259 \pm 0.023) \times 10^{-3}
$$

\noindent
show their equality within errors, indicating that $| \varepsilon'|
< < | \varepsilon|$. This is a consequence of the approximate
$\Delta I = 1/2$ rule.

In order to extract the corresponding phases $\phi_{+ -} , \phi_{00}$, 
one needs a mechanism of $K_L - K_s$ interference. This is possible
from either (i) Coherent Regeneration of $K_s$ from $K_L$, or
(ii) The Time-Distribution of $K \rightarrow \pi \pi$ events as
obtained from a pure flavour (strangeness) state prepared at $t = 0$.

The first method (i) needs the preparation of $K_L$-beams. The second
method (ii) makes use of flavour tag: either $K^0$
or $\bar{K}^0$. From the results on $K_L \rightarrow 2 \pi$, one
reproduces $Re \, \varepsilon_K \simeq 1.63 \times 10^{-3}$, in good
agreement with the value extracted from the charge asymmetry $\delta$
in semileptonic decays. The second method (ii) has been put into
practice by the CP-LEAR Collaboration at CERN recently 
\cite{PPV},
and their results for $\pi^+ \pi^-$ Decays are shown in Figure 2.

\mxfigura{9cm}{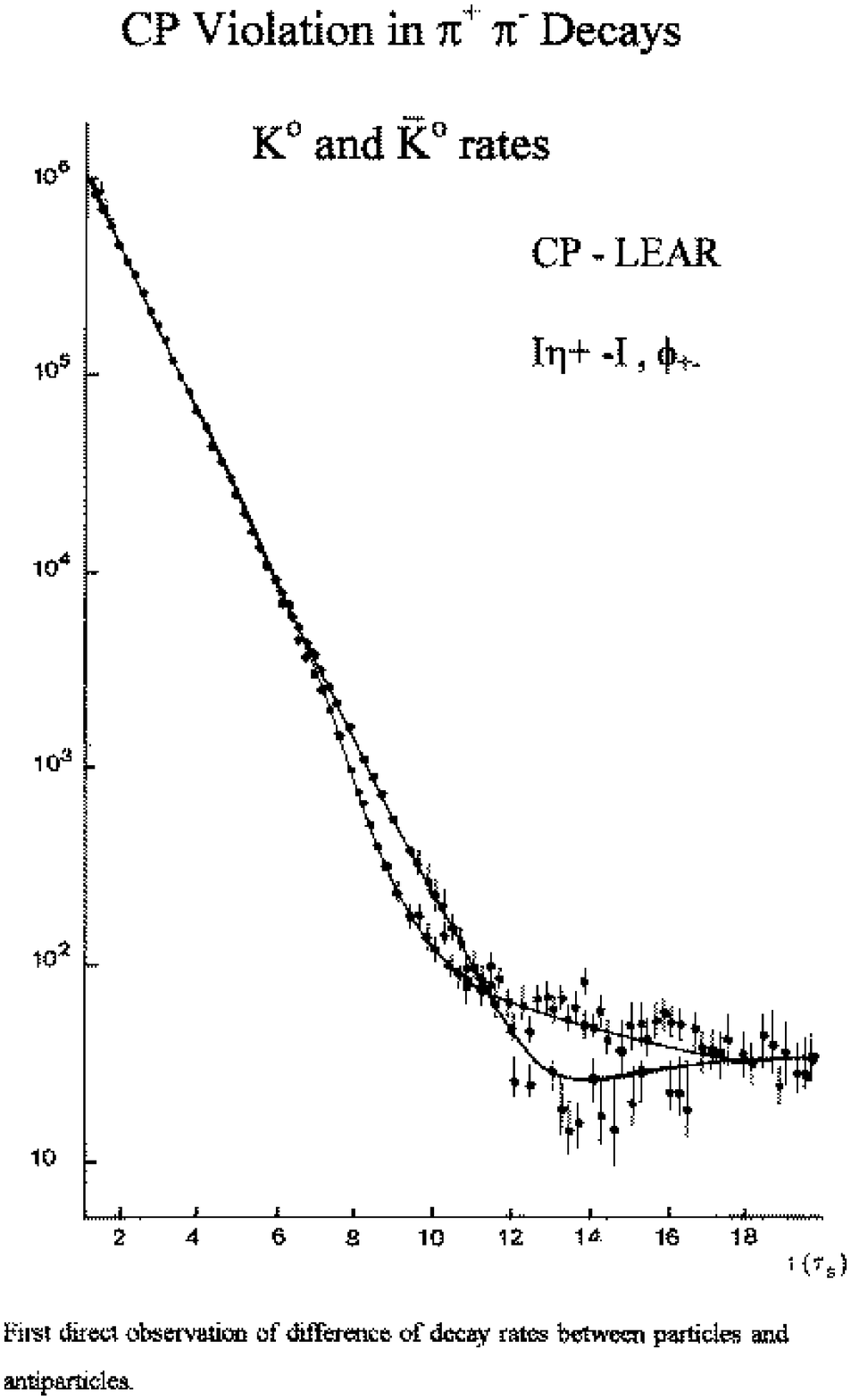}{Time Evolution of $K^0$ and $\bar{K}^0$ 
rates at CP-LEAR}{figura2}

These results represent the first direct observation of a difference
in the decay rates between particles and antiparticles.

The ratio $\varepsilon' / \varepsilon$ can be determined by the "method
of ratio of ratios" when comparing the $\pi^0 \pi^0$ and $\pi^+
\pi^-$ decay channels:

$$
Re (\frac{\varepsilon'}{\varepsilon}) \simeq \frac{1}{6}
\left\{ 1 - | \frac{\eta_{00}}{\eta_{+ -}}|^2 \right\}
$$

Present results give ($23.0 \pm 6.5) \times 10^{-4}$ for the
NA31- Collaboration \cite{GDB} at CERN and ($7.4 \pm 5.9) \times 10^{-4}$ for
the E731-Collaboration \cite{LKG}
at FermiLab. These values are statistically
compatible within 8$\%$. New experiments with better sensitivity
at CERN, FermiLab and a dedicated $\Phi$-Factory at Frascatti will push
the precision to better values. The Goal is to reach sensitivities
better than $10^{-4}$ !

\subsection{CKM Quark Mixing Matrix}

The standard mechanism to incorporate CP-violation in the present
electroweak theory is the Cabibbo-Kobayashi-Maskawa Complex Mixing
Matrix for 3 families of Quarks. It generates CP-violating effects in 
both:
(i) the $\Delta S = 2$ $\quad  K^0 \bar{K}^0$ transition, through the

\vspace{1cm}

\noindent
Box-Diagram

\mxfigura{7cm}{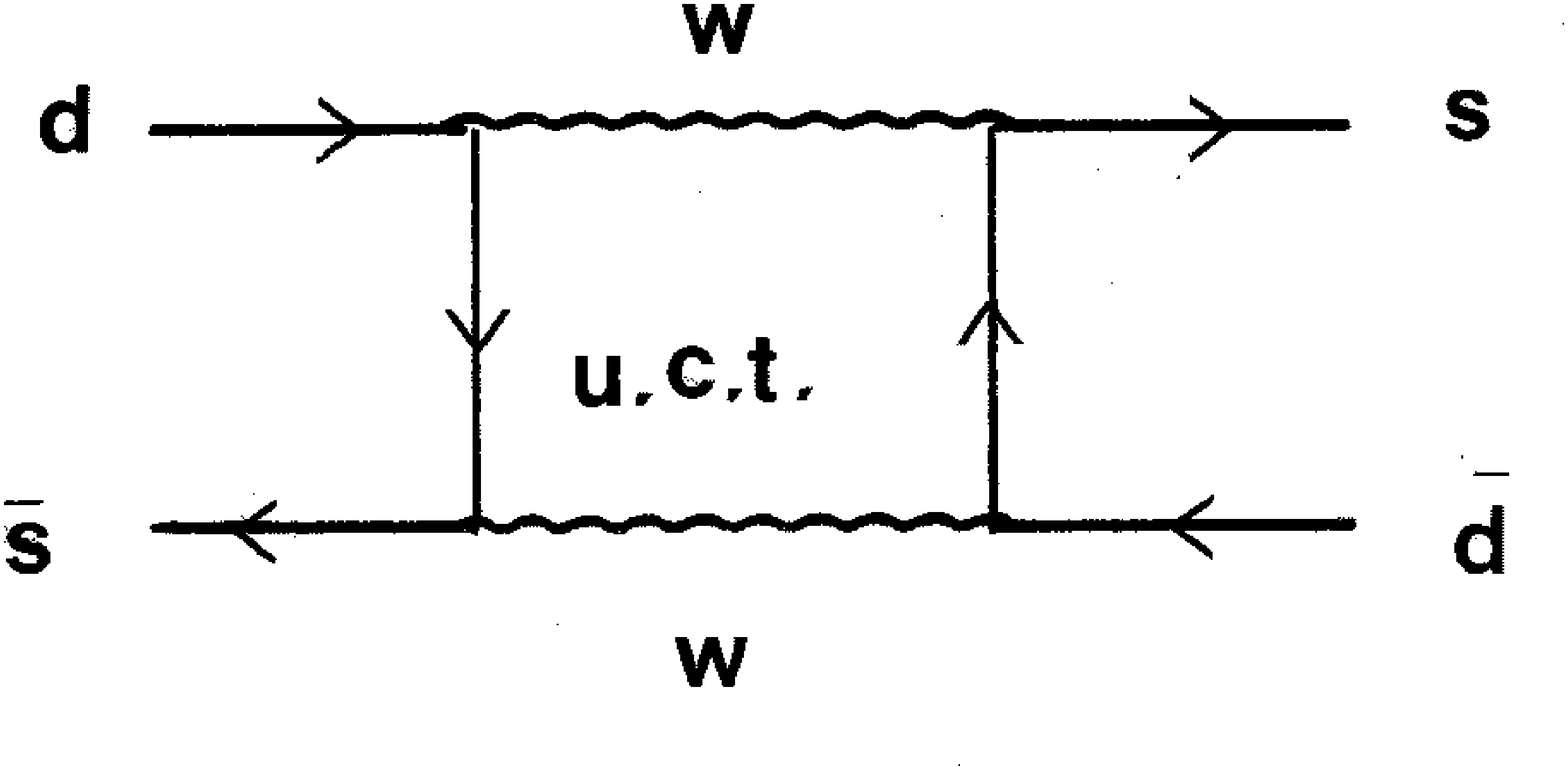}{Quark diagram for $\Delta S = 2 
\quad K^0 \protect\rightleftharpoons \bar{K}^0$}{figura3}

and (ii) the $\Delta S = 1$ Decay amplitude to non-strange quarks, 
through the

\vspace{1cm}

\noindent
Penguin-Diagram

\mxfigura{7cm}{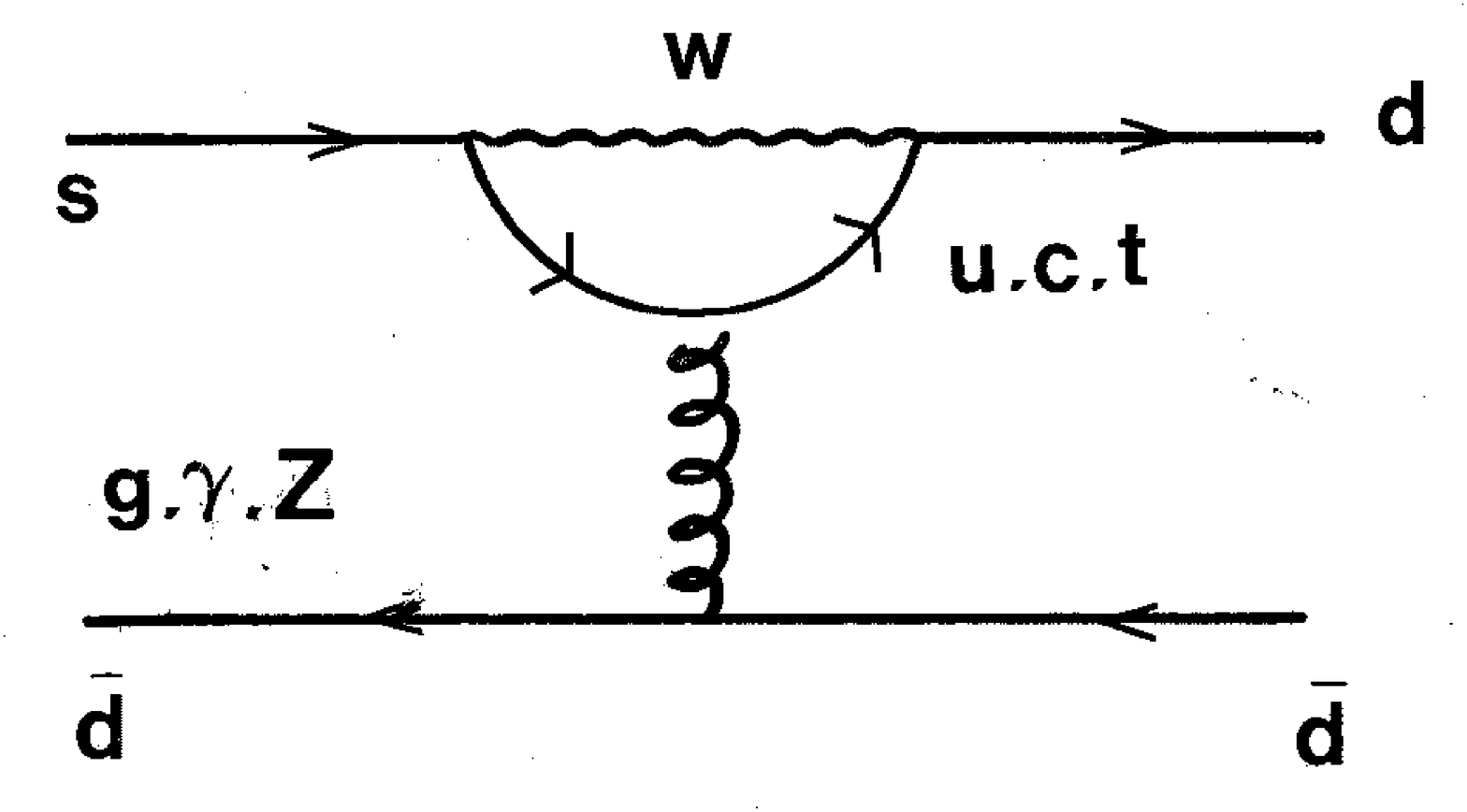}{Quark diagram for 
$\Delta S = 1 \quad K^0$ Decays}{figura4}

The real part of the $K^0 - \bar{K}^0$ mixing, which contributes to
$\Delta M$, is dominated by intermediate charm quarks running
 in the loop. Its main uncertainty comes from the hadronic
matrix element of the effective four-quark $\Delta S = 2$ operator,
when the higher degrees of freedom are integrated out. Its value is
parametrized by $B_K$, measuring it with respect to the 
vacuum-contribution in the intermediate state. 

The imaginary part of the box-diagram provides the explanation of
the indirect CP-violation $\varepsilon_K$ in the Standard Model and it is 
dominated by intermediate top quarks running in the loop. Its
experimental value can be analyzed in terms of the $V_{t d}$ matrix-element
of the CKM Mixing Matrix.

Using the Wolfenstein parametrization, the experimental value of
$\varepsilon_K$ specifies a hyperbola in the $(\rho, \eta)$
plane: the allowed region is shown in Fig. 5.

\mxfigura{10cm}{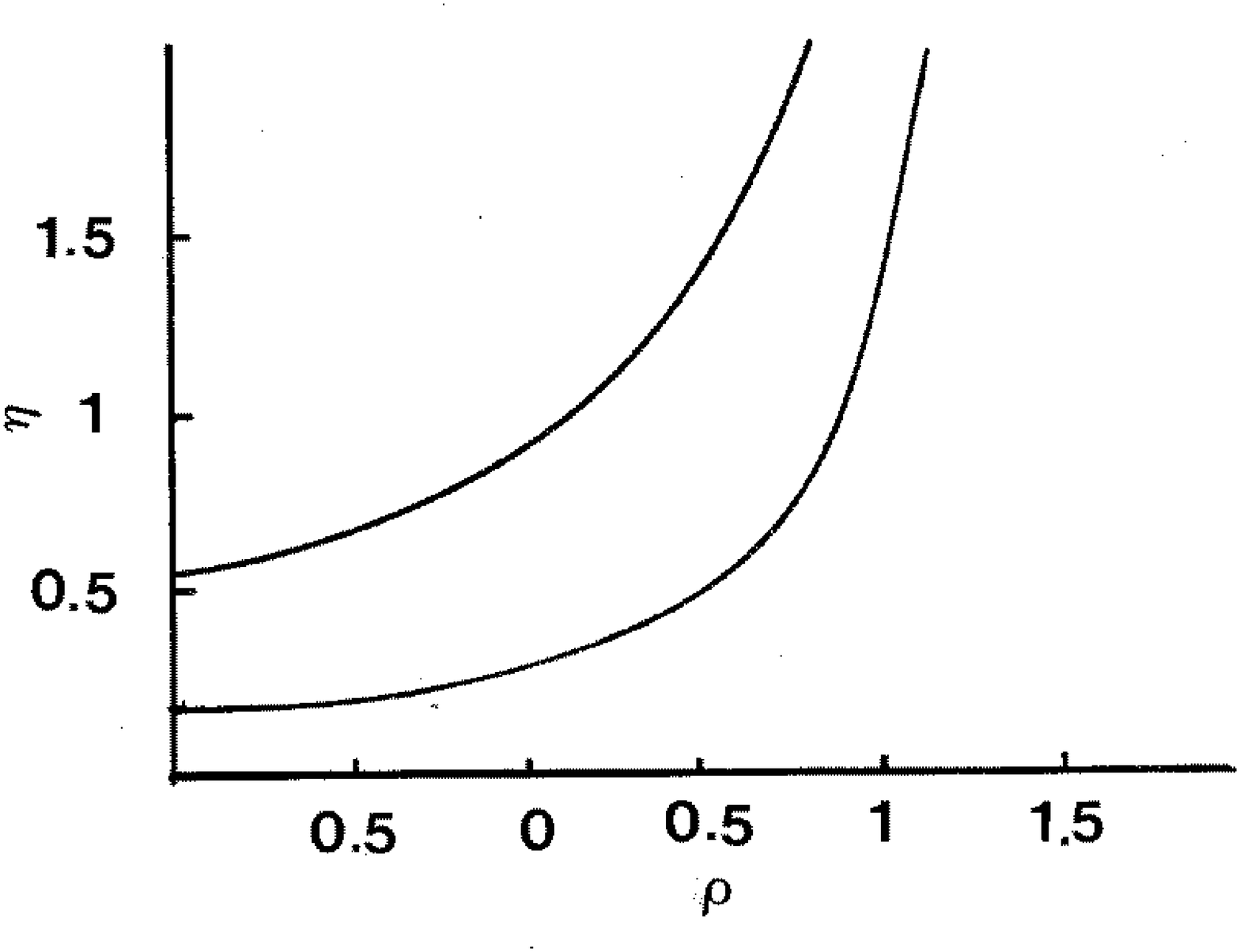}{Allowed region in the $(\rho, \eta)$ 
plane, constrained
from the experimental value of $\epsilon_K$.}{figura5}

A theoretical estimate of the direct CP-violation $\varepsilon' / 
\varepsilon$ is more involved. The quark diagrams giving $\Delta S = 1$
transitions, including the penguins, induce up to 10 four-quark 
operators once the higher degrees of freedom are integrated
out. A control of the long-distance effects becomes both fundamental
and difficult. Although the gluon-mediated penguin was initially
assumed to be the main ingredient, the $Z$-penguin contains
non-decoupling $m_t^2$ terms in the amplitude. The final result
is thus very sensitive to $m_t$. The present theory provides
values for $\varepsilon'/\varepsilon$ ranging from $- 3
\times 10^4$ to $10^{-3}$, depending on the hadronic physics being
taken by the Roma, M\"{u}nich or Dortmund groups.

\subsection{Coherent Decay of ($K^0 \bar{K}^0)$}

The two-particle $K^0 - \bar{K}^0$ system, such as is being produced
by $\Phi$-decay, has a total angular momentum $J = L$
and it is eigenstate of charge-conjugation $C$ with eigenvalue 
$C = (-1)^{L}$.

Bose statistics requires that the physical state has to be symmetric 
under the product of $C \times {\cal P}$, where $C$ is charge-conjugation
and ${\cal P}$ is permutation of space coordinates. In other
words, the physical state $| \psi >$ has to be eigenstate of $ C {\cal P}$
with eigenvalue +1.

Thus, as a consequence of Quantum Mechanics indistinguishibility, we
can write

$$
\begin{array}{ll}
| \psi > & = \frac{1}{\sqrt{2}} (1 + C {\cal P}) | LM ; K^0 \bar{K}^0 >
\\[2ex]
& = \frac{1}{\sqrt{2}} \left\{ | K^0  (\vec{k}) \bar{K}^0
(- \vec{k}) + (-1)^L | \bar{K}^0 (\vec{k}) K^0 (-\vec{k}) > \right\}
\end{array}
$$

If we impose the conditions of the $e^+ e^-$ machine at the $\Phi$-peak
(like DA$\Phi$NE at Frascati), we have $L = 1$ and $C = -$.
The produced state of $(K^0 - \bar{K}^0)$, when written in terms
of the physical neutral kaon states with definite mass and lifetime, is

$$
| K^0 \bar{K}^0 (C = -) > = \frac{1 + |\alpha|^2}{2 \sqrt{2} \alpha}
\left\{ | K_L (\vec{k}) K_s (- \vec{k}) > - | K_s (\vec{k}) K_L
(- \vec{k}) > \right\}
$$

\noindent
with no component of $K_s K_s$ or $K_L K_L$ states. 

We conclude \cite{JBE} :

\vspace{0.2cm}

\begin{tabular}{|l|}
\hline\\
(i) Bose statistics for $L = 1$ says that the state of two identical bosons
is \\
forbidden !\\
(ii) The result (i) applies to any time in the evolution of the system,\\
and thus we have simultaneously NO  $K^0 K^0$, NO $\bar{K}^0
\bar{K}^0$ !\\
(iii) The preparation of $(K^0  \bar{K}^0)$ from a $\Phi$-factory leads
to correlated\\
 (simultaneous) beams of $K_L$ and $K_s$ ! \\
\hline
\end{tabular}

\vspace{0.2cm}

The parameter $\alpha$ in the normalization of the state is the ratio
of the amplitudes

$$
\alpha \equiv \frac{p}{q} = \frac{1 + \varepsilon_K}{1 -
\varepsilon_K}
$$

By studying the decay amplitude from this correlated system, we ask
for the appearance of $X_1$ at time $t_1$ and $X_2$ at time $t_2$, as 
illustrated in Figure 6

\mxfigura{7cm}{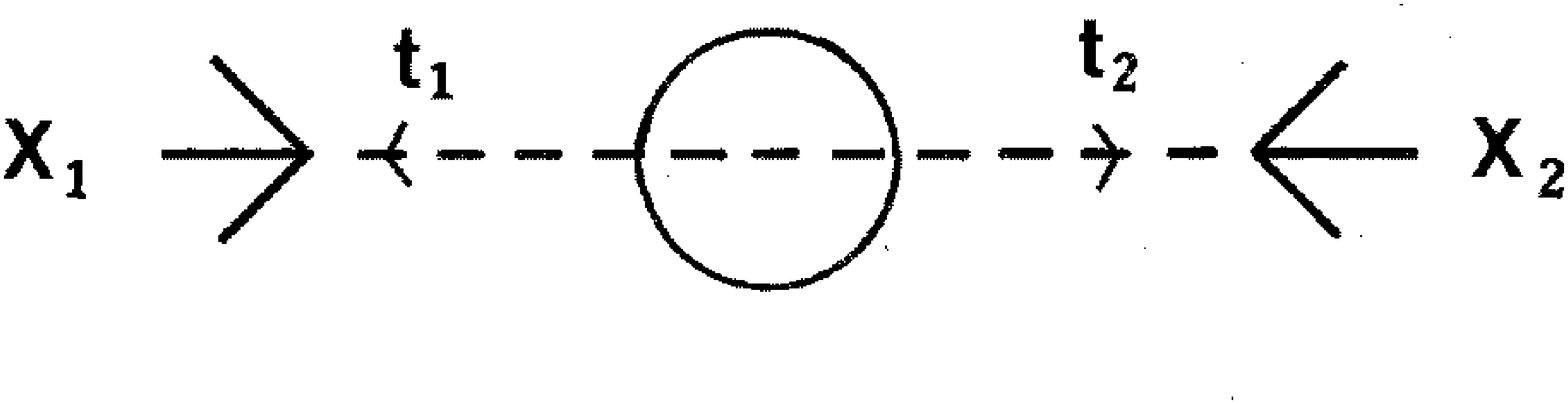}{The Decay of the Correlated 
$(K^0 \bar{K}^0)$}{figura6}

$$
\begin{array}{ll}
A (X_1, X_2) = \frac{1 + |\alpha|^2}{
2 \sqrt{2} \alpha}&   < X_1 | K_s > < X_2 | K_s > e^{-i
(\lambda_L + \lambda_s) t/2}\\[2ex]
& \left\{ \eta_1 e^{-i \Delta \lambda \Delta t/2} - \eta_2 e^{i \Delta
\lambda \Delta t/2} \right\}
\end{array}
$$

\noindent
where

$$
\eta_j = \frac{< X_j | K_L>}{< X_j | K_s >} \quad ; \quad
j = 1,2 
$$

\noindent
and

$$
\begin{array}{ccc}
\lambda_{s,L} = M_{s,L} - \frac{i}{2} \Gamma_{s,L} & ; & \Delta \lambda
= \lambda_s - \lambda_L \, ;\\
t = t_1 + t_2 & ; & \Delta t = t_2 - t_1
\end{array}
$$

Suppose that we detect the two decay products $X_1$ and 
$X_2$ at equal times  $t_1 = t_2$. For $\Delta t = 0$,
the decay amplutude gives

$$
A (X_1, X_2  \, ; \Delta t = 0) \propto (\eta_1 - \eta_2)
$$

In particular, for $X_1 = X_2$ one finds

$$
A [X_1 (t_1), X_1 (t_1)] = 0
$$

\noindent
which is a quantum mechanical correlation \cite{LIP} of the
Einstein-Podolsky-Rosen type implied by Bose
statisties.

If we select the two different $2 \pi$ decay channels of the neutral
kaon, i.e. $X_1 \equiv \pi^+ \pi^-$ and $X_2 \equiv \pi^0 \pi^0$, we
find at equal times

$$
A [\pi^+ \pi^- (t_1) , \pi^0 \pi^0 (t_1)] \propto (\eta_{+ -} - \eta_{00})
= 3 \varepsilon'
$$

This result has been discussed \cite{DUN} in the context of 
$\Phi$-factories,
trying to maximize the $\varepsilon'$ effect by comparing the observables
at different time slices. The intensity asymmetry, for a given
$\Delta t$, under the change $\Delta t \rightarrow - \Delta t$, is
very sensitive to $\varepsilon' /\varepsilon$ when $\Delta t$ is of the
order of $\tau_s$, the $K_s$ lifetime.

\subsection{Time Integrated Rates}

Intead of comparing the rates at different time intervals, for a given
decay channel $\pi^+ \pi^- \pi^0 \pi^0$ of the $K^0 - \bar{K}^0$
system, one can adopt a different strategy: the comparison of different
decay chanels in time integrated rates \cite{JBR}.

In particular,

$$
\frac{Br (\pi^+ \pi^- , \pi^+ \pi^-)}{Br (\pi^0 \pi^0, \pi^0 \pi^0)} =
\left[ \frac{Br (K_s \rightarrow \pi^+ \pi^-)}{
Br (K_s \rightarrow \pi^0 \pi^0)} \right]^2 
\left\{ 1 + 6 Re (\frac{\varepsilon' }{\varepsilon}) \right\}
$$

\noindent
and similarly \cite{JBR} for the other possible ratios.

\section{CP- Violation and $B$-Physics}

\subsection{Physics Motivation}

The main question is whether the origin of CP-violation can be explained
within the Standard Model or it needs physics  beyond the Standard Model.
Inside the Standard Model, the analysis for Flavour Mixing starts
from a Charged Current Lagrangian, with the Quark Mixing Matrix
described by 4 independent parameters. In the Wolfenstein
parametrization, for example, these are $[\lambda, A, \rho, \eta]$.

It has to be emphasized  that these are fundamental couplings
of the Standard Model Lagrangian, so their determination is of primordial
importance. In order to answer the question on the possible need of
beyond-the-standard-model physics, one has to look for ways to
overdetermine these parameters and show the internal consistency
of the model.

From the Unitary Quark Mixing Matrix

$$
V_{CKM} = \left(
\begin{array}{ccc}
V_{ud} & V_{us} & V_{ub}\\
V_{cd} & V_{cs} & V_{cb}\\
V_{td} & V_{ts} & V_{tb}
\end{array} \right) \; = \; 
\left[ \begin{array}{ccc}
 1 - \frac{\lambda^2}{2} & \lambda & A \lambda^3 (\rho - i \eta)\\
- \lambda &  1 - \frac{\lambda^2}{2} & A \lambda^2\\
A \lambda^3 (1 - \rho - i \eta) & - A \lambda^2 & 1 
\end{array}
\right] 
$$

\noindent
one can  establish 6 off-diagonal Unitarity Relations which,
represented in the complex plane, can be visualized (for 3 families) by 6 
Unitarity Triangles. One of these, that built from the
$d$- and $b$- columns, is written as

$$
(b, d) \rightarrow V_{ub}^* V_{ud} +
V_{cb}^* V_{cd} + V_{tb}^* V_{td}
= 0
$$

\noindent
and it has the peculiarity that the three sides are of the same
order of magnitude: $O  (\lambda^3)$.

In the Standard Model, a single phase parameter ($\eta$) is the only
possible source of CP violation, so that the predictions for CP-violating
phenomena are quite constrained. Moreover, the CKM mechanism requires
several necessary conditions in order to generate
an observable CP-violating effect: (i) All three families
are required to play and active role.
With only two fermion families, the Quark Mixing mechanism cannot
give rise to CP-violation. In the kaon system, for instance, CP-violation
effects can only appear at the one-loop level, where the top quark is 
present.
In B-physics, however, the decays generated by tree level diagrams 
can induce CP-violation effects; (ii) The quarks of a given charge
must be non-degenerate in mass. With degeneracy, the physical quark states
could be redefined in order to vanish the quark mixing matrix
element; (iii) all CKM-matrix elements must be non-zero. If any of
these conditions were not satisfied, the CKM-phase could be rotated away
by a redefinition of the quark fields. CP-violation effects are then
necessarily proportional to the product of all CKM angles.

All these necessary conditions are summarized in a single requirement
\cite{JRL} on the original quark-mass matrices $M_u$ and
$M_d$: 

$$ 
CP-\hbox{violation}  \Longleftrightarrow Im \{ det [M_u M_u^+,
M_d M_d^+ ] \} \neq 0
$$

In the case of 3 families, there is a unique combination of angles
and phases to generate CP-violation for all flavours that can be
considered:

$$
J \simeq A^2 \lambda^6 \eta \leq 10^{-4}
$$

Any CP-violation observable involves the product $J$ and, thus,
violations of the CP symmetry are necessarily small.

One can make some general statements:

(i) In order to generate sizeable CP-violating asymmetries
$[(\Gamma - \bar{\Gamma})/(\Gamma + \bar{\Gamma})]$ one should look for
very suppressed decays, where the decay widths already involve
small CKM matrix elements.

(ii) In the Standard Model, CP violation is a low-energy phenomenon, in
the sense that any effect should disappear with $\frac{m_c^2 - m_u^2}{
s} < < 1$.

(iii) $B$ decays are the optimal place for CP-violation signals to show 
up. They involve small CKM matrix elements in the decays
and are the lowest mass processes where the three quark families play
a tree-level role.

The (bd) Unitarity Triangle is shown in Fig. 7, where it has
been scaled by dividing its sides by $|V_{cb}^* V_{cd}|$. In the
Wolfenstein parametrization, this is real. Aligning it, with length equal
to 1, along the real axis, the coordinates of the 3 vertices are then
$(0, 0), (1,0)$ and $(\rho, \eta)$. Note that, although the orientation
of the triangle in the complex plane is phase-convention dependent, the 
triangle itself is a physical object: the length of the sides and the
angles can be directly measured.

\mxfigura{9cm}{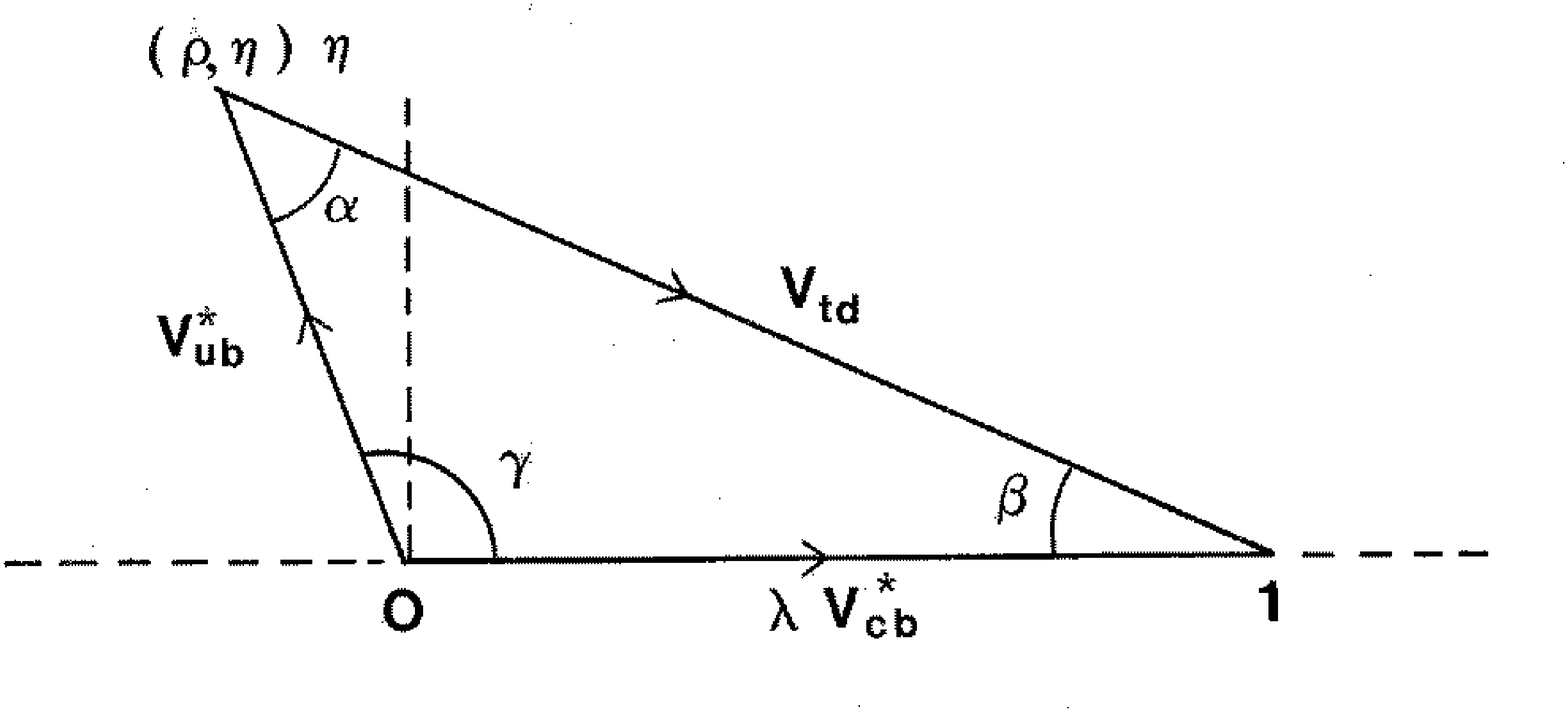}{The $(bd)$ Unitarity Triangle}{figura7}

The sides of the Unitarity Triangle are determined from the measured
ratio $\Gamma (b \rightarrow u)/ \Gamma (b \rightarrow c)$ and from
$B_d^0 - \bar{B}_d^0$ mixing:

$$
R_b = \left| \frac{V_{ub}^* V_{ud}}{V_{cb}^* V_{cd}} \right| \approx
\left| \frac{V_{ub}}{\lambda V_{cb}} \right| \approx  \sqrt{\rho^2 + \eta^2}
$$

$$
R_t = \left| \frac{V_{tb}^* V_{td}}{V_{cb}^* V_{cd}} \right| \approx 
\left|
\frac{V_{td}}{\lambda V_{cb}} \right| \approx  \sqrt{(1 - \rho)^2 +\eta^2}
$$

The significance of the $V_{td}$ coupling is seen in the box diagram of
Figure 8, relevant to the $B^0_d - \bar{B}^0_d$ mixing.
A priori, the measurement of the two sides, performed through CP-conserving
observables, could make possible to establish that the area is not
vanishing and that Fig. 7 is indeed a triangle. With the present
theoretical and experimental uncertainties, this is however not
possible.

\mxfigura{3cm}{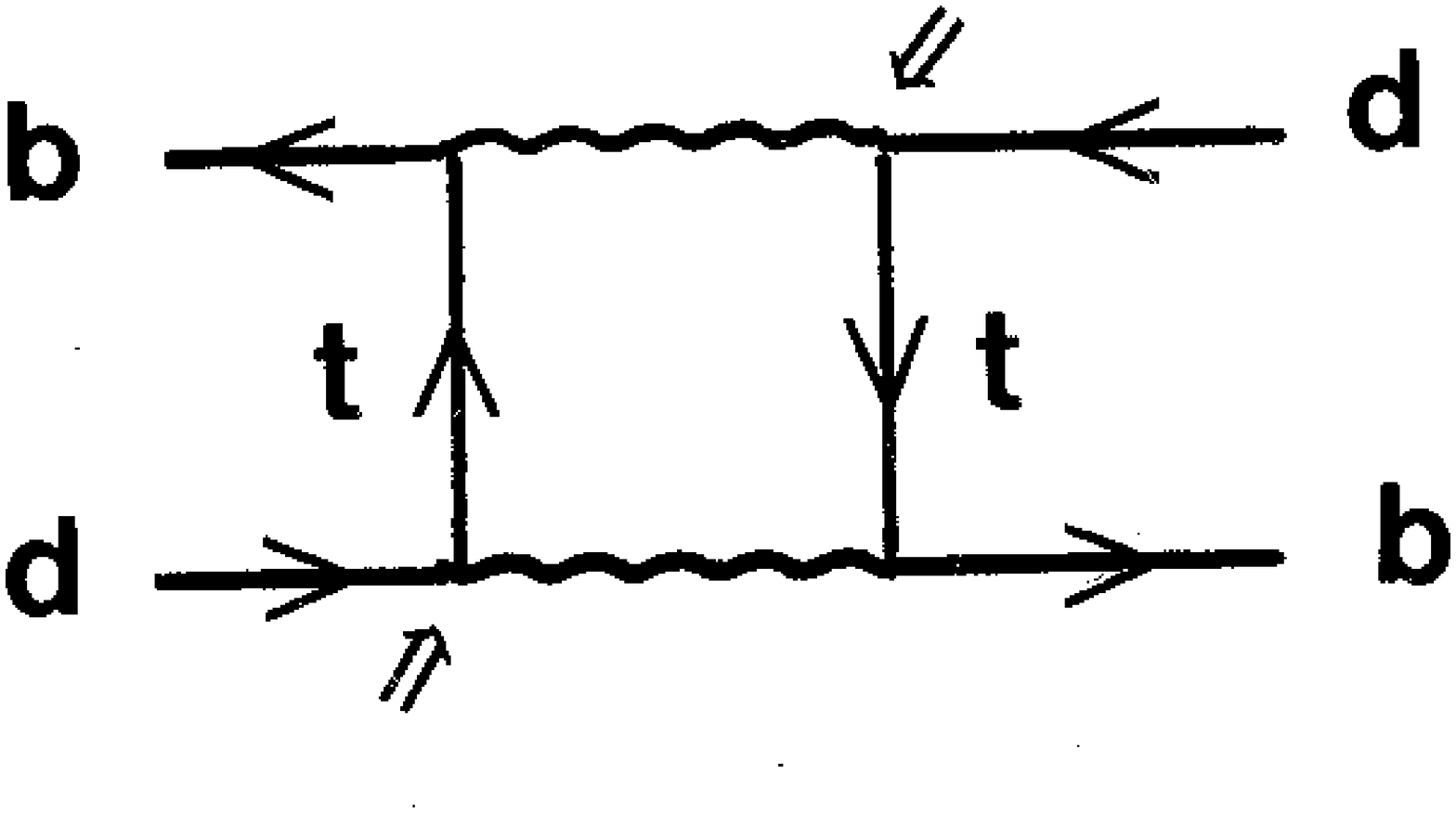}{Quark diagram responsible of the $B^0_d 
- \bar{B}_d^0$
mixing. The arrows indicate the $V_{td}$ coupling.}{figura8}

In section 2 we have seen that the experimental value of $\epsilon_K$
specifies a hyperbola in the ($\rho, \eta)$ plane. This constraint,
together with the constraints from  $R_b$ and $R_t$, which result
in the circles centered at ($0,0)$ and $(1,0)$, respectively, give an allowed
range of values \cite{AAL} for ($\rho, \eta)$ as shown in Fig. 9.
The final value is given by the intersection of all constraints.

\mxfigura{7cm}{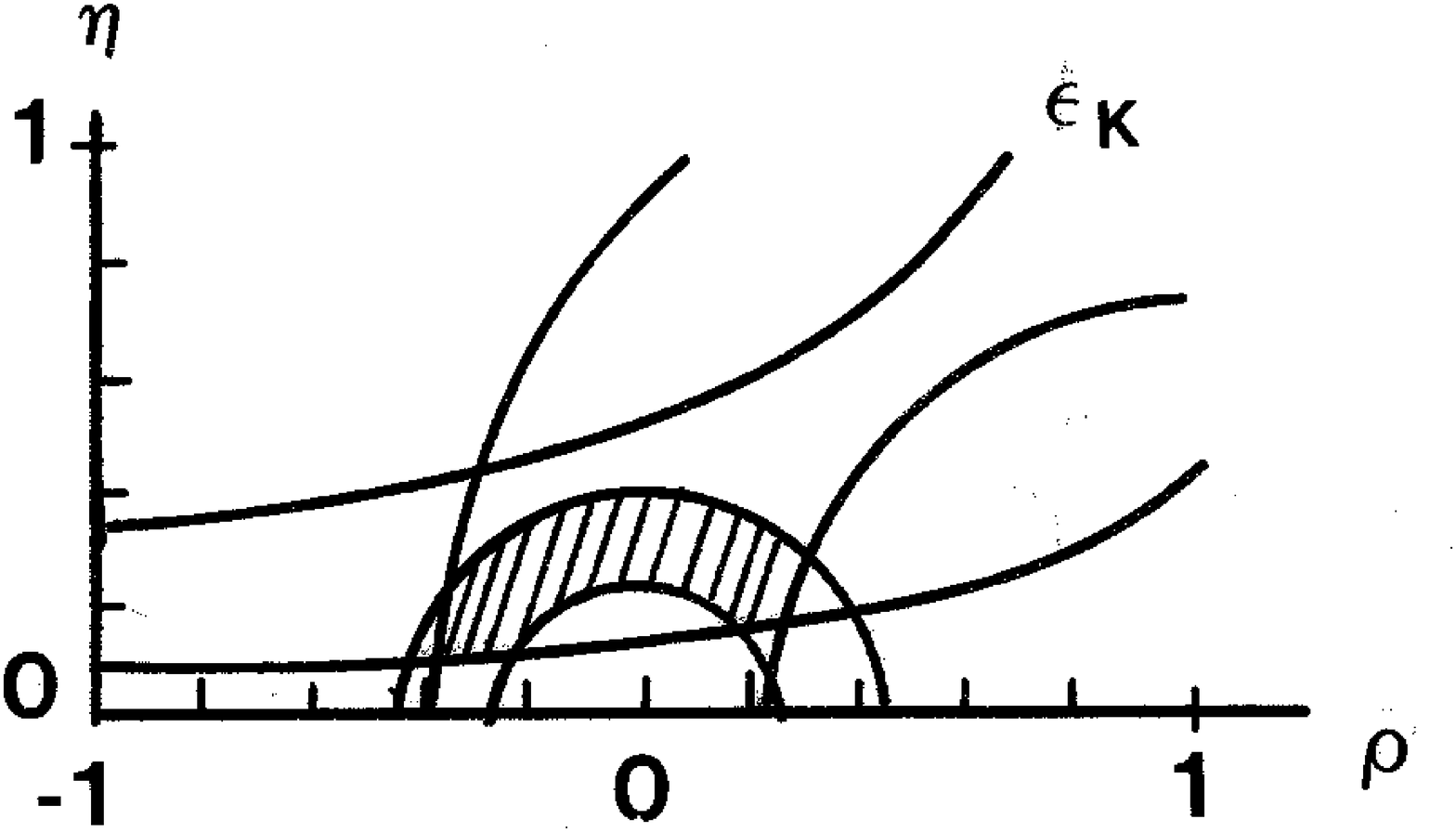}{Constraints for $(\rho, \eta)$ coming 
from the measured
values of $R_b, R_t$ and $\epsilon_K$.}{figura9}

\subsection{Principle of the $CP$-violation measurement}

i) With a large $B^0 - \bar{B}^0$ mixing we can generate a CP-violating
interference in the following way. There are quite a few non-leptonic
final states which are reachable both from a $B^0$ and a $\bar{B}^0$.
For these flavour non-specific decays the $B^0$ (or $\bar{B}^0)$
can decay directly to a given final state $f$, or do it after
the meson has been changed to its antiparticle via the mixing process;
i.e., there are two different amplitudes, $A (B^0 \rightarrow f)$
and $A (B^0 \rightarrow \bar{B}^0 \rightarrow f$), corresponding to two
possible decay paths, as shown in Fig. 10.

\mxfigura{7cm}{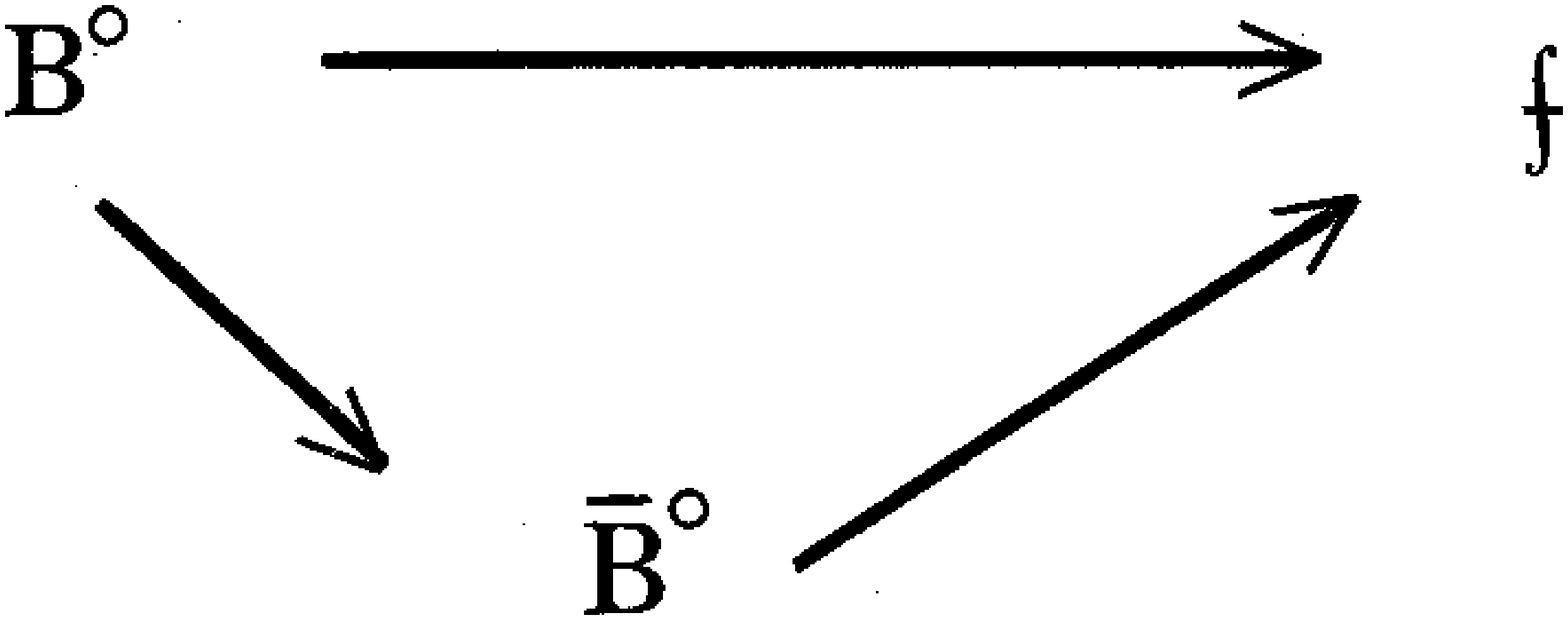}{Two decay paths 
from $B^0$ to $f$, with interfering
amplitudes.}{figura10}

\noindent
CP-violating effects can then result \cite{ACA} from the interference
of these two amplitudes. To build the associated asymmetry one needs
to TAG the initial B-flavour.

The time-dependent decay probabilities for the decay of a neutral $B$
meson created at the time $t_o = 0$ as a pure $B^0 (\bar{B}^0)$ into the
final state $f (\bar{f} \equiv CP \, f)$ are (we neglect the
tiny $\Delta \Gamma_{B^0}$ corrections):

$$
\begin{array}{ll}
\Gamma [B^0 (t) &  \rightarrow f] \sim \frac{1}{2} e^{- \Gamma t} 
|A_f|^2 \left\{ [1 + | \bar{\rho}_f|^2] +  \right.  \\
&  \left. 
+ [1 - |\bar{\rho}_f|^2 ] \cos (\Delta M t) - 2 Im (\frac{q}{p} \bar{\rho}_f)
\sin (\Delta M t) \right\}
\end{array}
$$

$$
\begin{array}{ll}
\Gamma [\bar{B}^0 (t) &  \rightarrow \bar{f}] \sim \frac{1}{2} e^{- \Gamma t} 
|\bar{A}_{\bar{f}} |^2 \left\{ [1 + | \rho_{\bar{f}} |^2] +  \right.  \\
&  \left. 
+ [1 - |\rho_{\bar{f}}|^2 ] \cos (\Delta M t) - 2 Im (\frac{p}{q} 
\rho_{\bar{f}}) \sin (\Delta M t) \right\}
\end{array}
$$

\noindent
where we have introduced the notation

$$
\begin{array}{l}
A_f \equiv A [B^0 \rightarrow f], \quad \bar{A}_f \equiv
- A [\bar{B}^0 \rightarrow f], \quad \bar{\rho}_f \equiv \bar{A}_f/A_f\\
A_{\bar{f}} \equiv A [B^0 \rightarrow \bar{f}], \quad \bar{A}_{\bar{f}}
 \equiv
- A [\bar{B}^0 \rightarrow \bar{f}], \quad \rho_{\bar{f}}
 \equiv A_{\bar{f}}/\bar{A}_{\bar{f}}\\
\end{array}
$$

CP-invariance demands the probabilities of CP conjugate processes to be
identical. Thus, CP conservation requires $A_f = \bar{A}_{\bar{f}} , 
A_{\bar{f}} = \bar{A}_{f}, \bar{\rho}_f = \rho_{\bar{f}}$ and
$Im (\frac{q}{p} \bar{\rho}_f) = Im (\frac{p}{q} \rho_{\bar{f}})$.
The violation of any of the first three equalities would be a signal
of direct CP-violation, i.e., in the decay amplitudes. The fourth
equality tests CP violation generated by the interference of the
direct decay $B^0 \rightarrow f$ and the mixing-induced decay
$B^0 \rightarrow \bar{B}^0 \rightarrow f$, as shown in Fig. 10.

If the final state is a CP-eigenstate, $\bar{f} = \tau_f f$
with $\tau_f = \pm 1$ and there is only one
transition decay amplitude \footnote{Each transition amplitude
is identified as the weak interaction matrix element of a given operator
with a definite CP-phase}, we have 

$$
\rho_{\bar{f}} = \bar{\rho}_f = \tau_f \, e^{2 i \phi_{WD}}
$$

$$
\frac{\Gamma [B^0 (t) \rightarrow f ] - \Gamma[\bar{B}^0 (t) \rightarrow f]}
{\Gamma [B^0 (t) \rightarrow f] - \Gamma[\bar{B}^0 (t) \rightarrow f]}
= \tau_f \sin (2 \Phi) \sin (\Delta M t)
$$

\noindent
where $\Phi$  is the convention-independent CP-phase which results
from the combination of the phase $\phi_{WD}$ of the decay amplitude
and the phase of  the mixing parameter $\frac{p}{q}$. This mechanism
is thus called "the interplay between Mixing and Direct CP-Violation".

ii) Consider now two interfering decay amplitudes which contribute to a
given process $B \rightarrow f$:

$$
\begin{array}{l}
A (B \rightarrow f) = M_1 e^{i \phi_1} e^{i \alpha_1} + M_2 e^{i \phi_2}
e^{i \alpha_2} \\
A (\bar{B} \rightarrow \bar{f}) = M_1 e^{- i \phi_1} e^{i \alpha_1} + 
M_2 e^{- i \phi_2} e^{i \alpha_2}
\end{array}
$$

\noindent
where $\phi_1, \phi_2$ are the corresponding weak decay phases and $\alpha_1,
\alpha_2$ are the final state "strong" interaction phases. They are
such that, in going to the CP-conjugated process, the weak phases
change their sign, whereas the strong phases keep their sign. The strong
phases originate from the beyond the Born approximation complexity
of the decay amplitudes. The rate Asymmetry is thus given by \cite{JBU}

$$
\frac{\Gamma (B \rightarrow f) - \Gamma (\bar{B} \rightarrow \bar{f})}
{\Gamma (B \rightarrow f) + \Gamma (\bar{B} \rightarrow \bar{f})} =
$$

$$
\frac{- 2 M_1 M_2 \sin (\phi_1 - \phi_2) \sin (\alpha_1 - \alpha_2)}
{|M_1|^2 + |M_2|^2 + 2M_1 M_2 \cos (\phi_1 - \phi_2) \cos
(\alpha_1 - \alpha_2)}
$$

\noindent
so that one needs \cite{JBU}:

1) Two, at least, interfering amplitudes.

2) Two different weak phases $\phi_1 \neq \phi_2$

3) Two different strong phases $\alpha_1 \neq \alpha_2$.

4) To enhance the asymmetry, $M_1$ and $M_2$ should be of
comparable size.

This "Direct CP-Violation" operates either for charged B's or neutral
B's with flavour specific decays.

iii) There is a third method to generate CP-Violation, that of 
CP-Violation in $B^0 - \bar{B}^0$ Mixing or "Indirect CP-Violation".
Take the semileptonic decays of a $B^0 - \bar{B}^0$ pair with equal
sign dileptons: $N ( l^+ l^+)$ or $N (l^- l^-)$ is a signal of mixing,
their difference ia a signal of CP-Violation in this mixing.

One gets the semileptonic asymmetry of $B^0 \rightarrow X l \nu_l$

$$
a_{SL} = \frac{N (l^+ l^+) - N (l^- l^-)}
{N (l^+ l^+) + N (l^- l^-)} = 
\frac{|p/q|^2 - |q/p|^2}{|p/q|^2 + |q/p|^2} \simeq 4 Re \, \bar{\varepsilon}_B
$$

The off-diagonal matrix elements of the  effective mixing matrix of
$B^0 - \bar{B}^0$ give rise to a difference of the masses and widths
of the two physical eigenstates. In the Standard Model the two
complex eigenvalues satisfy

$$
| \frac{\Delta \Gamma_B}{\Delta M_B}| \sim | \frac{\Gamma_{12}}{M_{12}}|
\sim \frac{m_b^2}{m_t^2}
$$

To generate $Re (\bar{\varepsilon}_B) \neq 0$ one needs both
$\Delta \Gamma_B \neq 0$ and a missalignment of the (complex) values
of $\Gamma_{12}$ and $M_{12}$. The relative argument between $M_{12}$
and $\Gamma_{12}$ is expected to be of $0 (m_c^2 /m_b^2)$ in the
Standard Model. All together, $\alpha_{SL}$ is expected to be less
than $10^{-3}$ for $B_d^0$ and less than $10^{-4}$ for $B^0_S$
 and thus outside of the capabilities of the next future experimental
facilities.

\subsection{Rate Asymmetries}
 
- By using the first method discussed in the last Section, one considers
the process \cite{BIA} $B \rightarrow J/\psi \, K_s$ able to measure sin
$(2 \beta)$, where $\beta$ is the angle of the unitarity
triangle shown in Fig. 7. The quark diagrams responsible of this process
are a tree diagram and a penguin diagram, as shown in Fig. 11

\mxfigura{9cm}{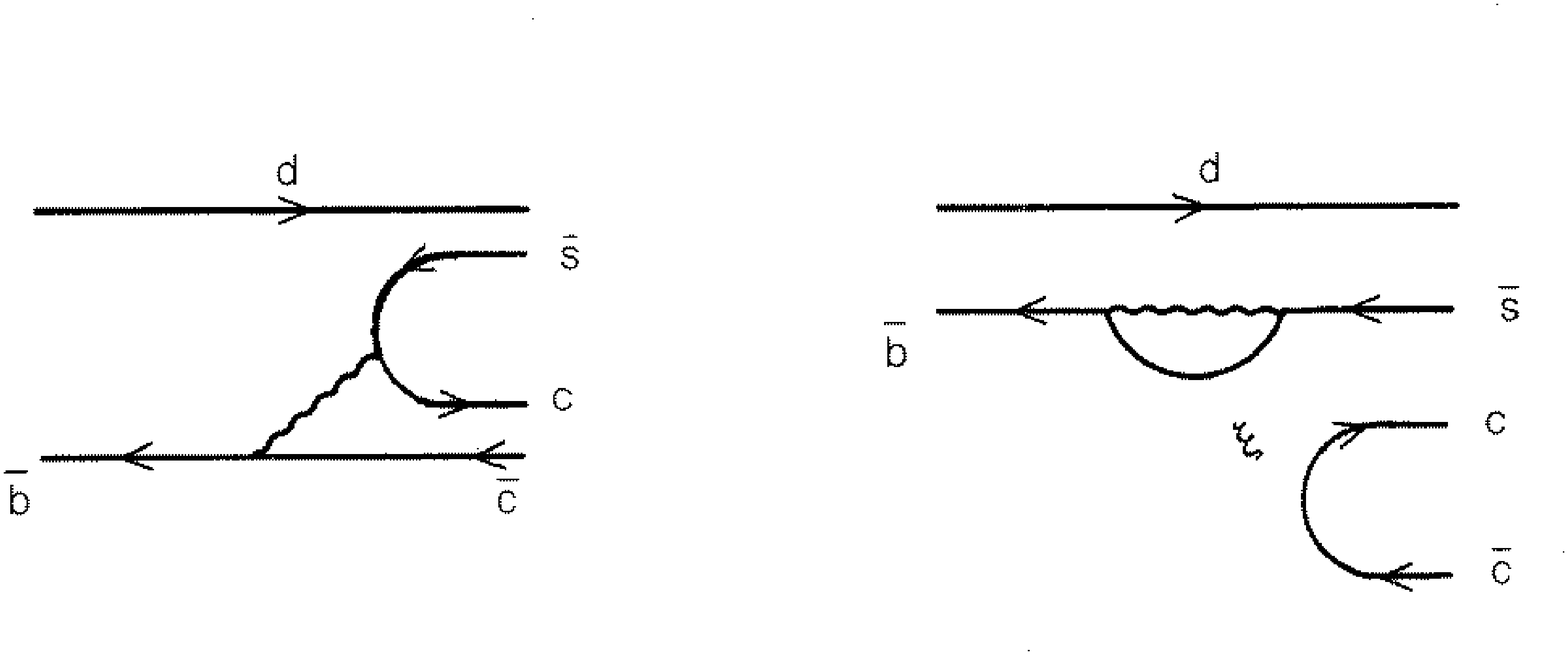}{Tree and penguin diagrams 
relevant to the process
$B \rightarrow J/\psi \, K_s$.}{figura11}

The penguin contribution is small, estimated to be less than
1$\%$ of the tree amplitude, and furthermore one discovers by inspection
that both diagrams contain the same CP-phase, i.e., there is
a unique (complex) hadronic amplitude to describe the exclusive 
process.

For $B'_d s$ in the initial state, one has $\Delta \Gamma/ \bar{\Gamma}
\sim 5 \times 10^{-3}$, so $\Delta \Gamma$ can be neglected in the time
distribution.
The corresponding rate asymmetry for this definite final state $f \equiv
J/\psi \, K_s$, eigenstate of CP, leads to 

$$
\frac{\Gamma_f - \bar{\Gamma}_f}{\Gamma_f + \bar{\Gamma}_f}
= - \sin (2 \beta) \, \sin (\Delta M t)
$$

\noindent
independent of the (unique) hadronic amplitude !

This asymmetry is expected to be detected and measured at the 
Tevatron, the
$B$-factories,
 HERA-B  and  at the LHC  experiments. Precision determinations
with $\sigma (\sin 2 \beta) \sim \, 0.02$ can be expected.

Taking into account that the $(c \bar{c})$ system in the quark diagrams
of Fig. 11  is, by itself, composed of charge-conjugated quarks,
one can envisage the consideration of the semiinclusive process
\cite{JBK} $B \rightarrow K_s X (c \bar{c})$. No cancellation of 
the rate asymmetries is expected when summing over the different
hadronic  states of the ($c \bar{c}$) spectrum.

- With the aim to extract the angle $\alpha$ of the unitarity triangle,
one considers the exclusive process $B_d \rightarrow \pi^+ \pi^-$,
where $f  \equiv \pi^+ \pi^-$ is again a CP-eigenstate.
The quark diagrams, shown in Fig. 12, contributing to the process
are again a tree diagram and a penguin diagram

\mxfigura{9cm}{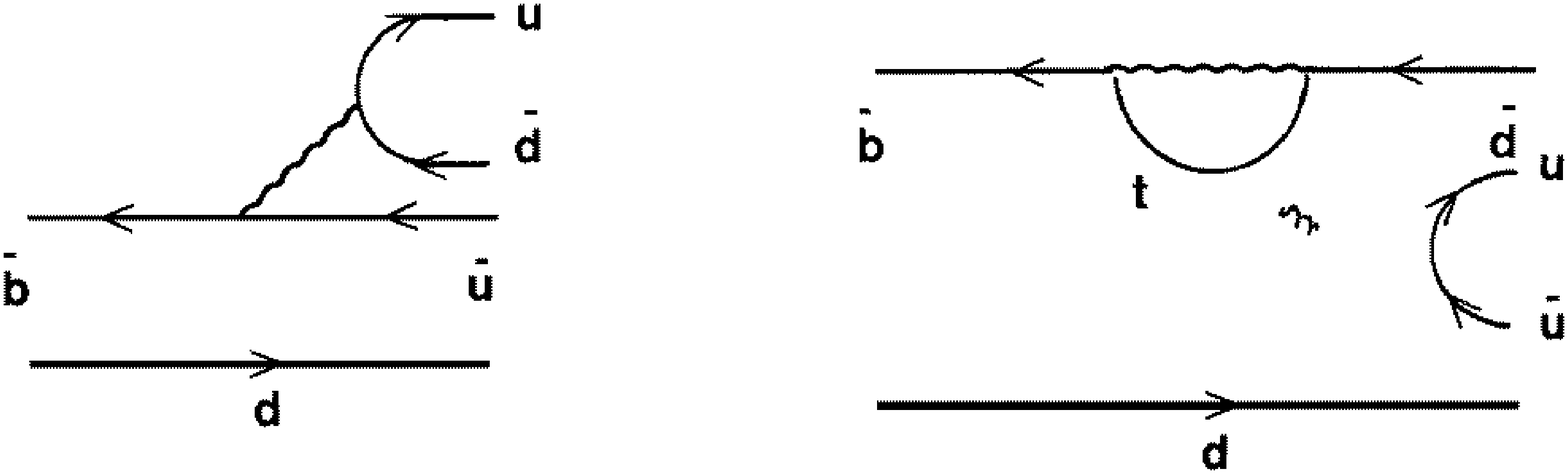}{Tree and penguin diagrams 
relevant to the process
$B_d \rightarrow \pi^+ \pi^-$.}{figura12}

The tree amplitude is governed by the quark mixing matrix element

$$
V^*_{ub} V_{ud} \sim A \lambda^3 (\rho + i \eta)
$$

\noindent
whereas the penguin amplitude contains

$$
V^*_{tb} V_{td} \sim A \lambda^3 (1 - \rho - i \eta)
$$

\noindent
so that both are of similar CKM strength. The effect of the loop and the
gluonic exchange in the penguin amplitude has been estimated to give
a relative contribution of the order of 10$\%$-20$\%$. An independent
measure of these two contributions and their relative magnitudes is
highly desirable.

Taking into account the contribution of the two different hadronic
amplitudes, one induces a Time-Dependent CP-violating Rate
Asymmetry

$$
\frac{N - \bar{N}}{N + \bar{N}} = \, a \cos (\Delta mt) + b \sin
(\Delta m t)
$$

\noindent
where "a" is obtained from the interference of the Direct CP violating
Decay Amplitudes, whereas "b" comes from the interplay between Mixing
and Decay.

Although the process $B_d \rightarrow \pi \pi$ is highly CKM-forbidden,
the main problem is not statistics, particularly in the hadronic
machines. Its major source of background comes from the accompanying
decays $B_d \rightarrow K^+ \pi^- \, ; B_s \rightarrow K^+ \pi^- \,
, K^+ K^-$, which by themselves could have contributions to the asymmetry
and not only to the rate.
The virtues of having particle identification in the detector are
evident, taking into account that: i) those backgrounds are reduced
by Particle Identification, ii) the amount of background and its shape
can be studied by reconstructing each decay mode using Particle Identification.
These virtues are illustrated in Fig. 13, as advanced by the LHC-B 
collaboration

\mxfigura{9cm}{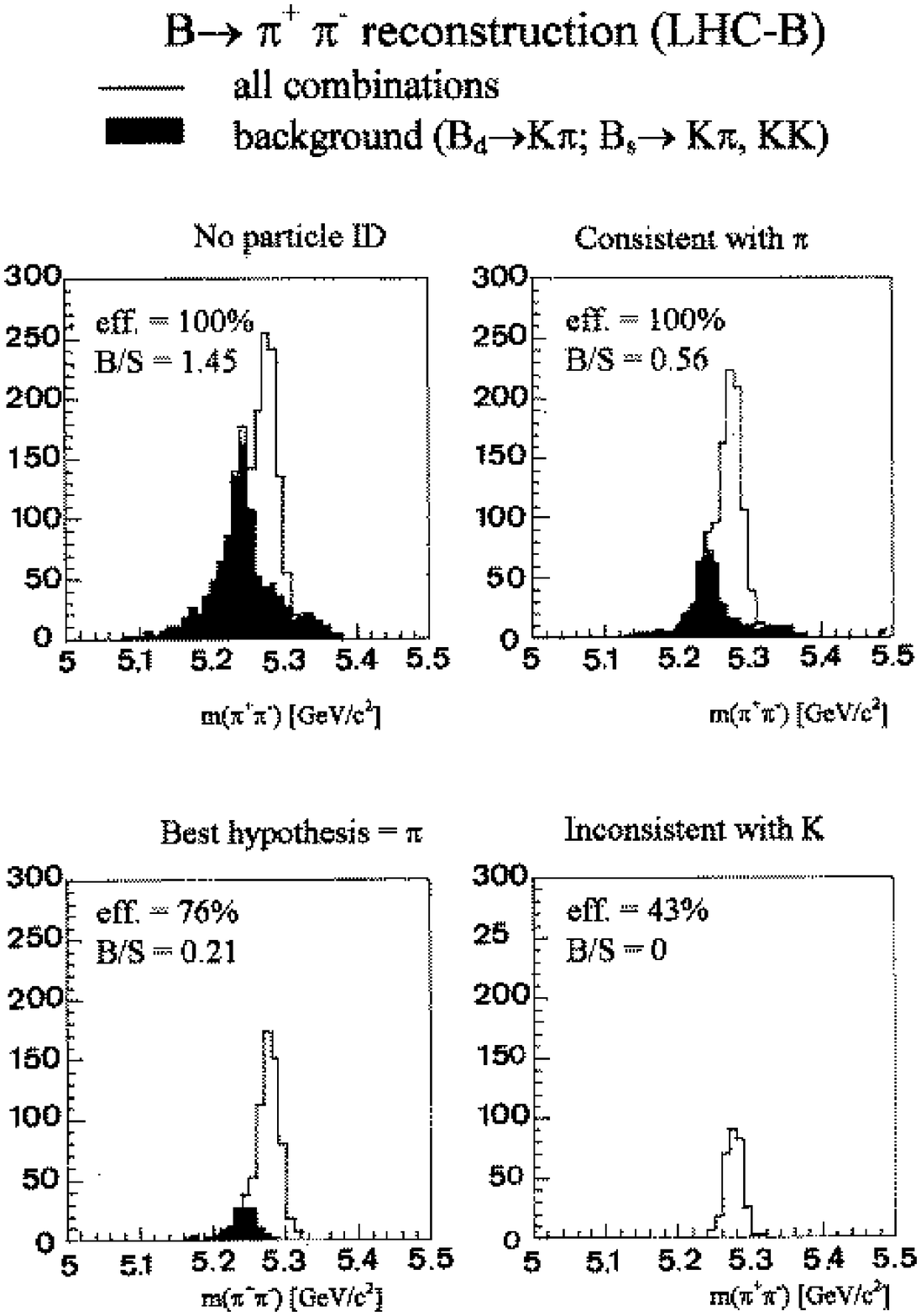}{The virtues of Particle Id for $B_d \rightarrow \pi
\pi$.}{figura13}

A purity of the data sample near 100$\%$ can be envisaged by reducing
the efficiency to about 50$\%$.

From the theoretical side, the absence of the penguin amplitude would
lead to $a = 0$, $b = - \sin (2 \alpha)$, where $\alpha$ as shown in
the unitarity triangle of Fig. 7. The process $B_d \rightarrow
\pi^+ \pi^-$ is, however, theoretically difficult, because the presence
of the penguin  amplitude $P$ affects the extraction of $\alpha$ as
obtained from the tree amplitude $T$. To leading order in $| P/T|$
small, one has

$$
\left\{
\begin{array}{c}
a = 2  | \frac{P}{T}| \sin (\delta_P - \delta_T) \sin \alpha\\
b = - \sin (2 \alpha) - 2 | \frac{P}{T}| \cos (\delta_P - \delta_T)
\sin \alpha \cos 2 \alpha
\end{array} \right.
$$

\noindent
where $\delta_{T (P)}$ is the elastic-final-state-interaction phase
shift for the $T (P)$ hadronic amplitude.

If $| P/T|$ were known independently, the value of $\alpha$
could be extracted from the measurement of both "a" and "b". The
quantity "b" is very sensitive to $\alpha$  and much less to the phase
shift difference $\delta_P - \delta_T$, whereas the quantity
"a" can determine $\delta_P - \delta_T$. The correlation between
$\alpha$ and $\delta_P - \delta_T$ is quite small.

To determine $|P/T|$ experimentally, Gronau and London have used 
the isospin triangle relations \cite{MGR} among the amplitudes
$M_{+ 0}$ for $B_u^{\pm} \rightarrow \pi^\pm \pi^0, M_{+ -} $ for
$B_d \rightarrow \pi^+ \pi^-$ and $M_{00}$ for $B_d \rightarrow
\pi^0 \pi^0$

$$
M_{+ -} \, + \sqrt{2} M_{00} = \sqrt{2} M _{+ 0}
$$

If \underline{all} the three decay modes are measured, the
value of $|P/T|$ can be extracted. However, $B_d \rightarrow \pi^+
\pi^-$  is colour allowed in the quark diagram, whereas $B_d
\rightarrow \pi^0 \pi^0$ is colour suppressed and it needs a colour
rearrangement. What is the price to be paid by this dynamics could be 
as high as
to lower the branching ratio for $\pi^0 \pi^0$ by a factor 10.

The channel $B_d \rightarrow \pi^+ \pi^-$ could have other theoretical
problems, like the existence of an electroweak (beyond the gluonic)
penguin  amplitude and/or inelastic-final-state-interaction effects. 
These problems need further scrutiny \cite{KRP}.

- With the strategy to show the consistency of the treatment, different
methods to extract the angle $\gamma = \pi - 
\alpha - \beta$ from rate-asymmetries have been
proposed. I mention here two   of these proposals \cite{RAL}:

i) Use of four Time-Dependent Decay Rates from the $B_s$-meson

$$
\begin{array}{l}
B_s \rightarrow D_s^- K^+ \stackrel{CP}{\longleftrightarrow}
\bar{B}_s \rightarrow D_s^+ K^-\\[2ex]
\bar{B}_s \rightarrow D_s^- K^+ \longleftrightarrow  B_s \rightarrow
D_s^+ K^- \, ,
\end{array}
$$

\noindent
because the final state is not a CP-eigenstate in this case.

The Proper Time Distributions depend on the parameters $x_s, \Delta
$ and $\gamma$, describing the mixing in the $B_s$-system, the 
final-state-strong-interaction phase and the relative weak CP-phase.
The tree diagrams from $B_s$ and $\bar{B}_s$ to a given final
state are shown in the Fig. 14

\mxfigura{9cm}{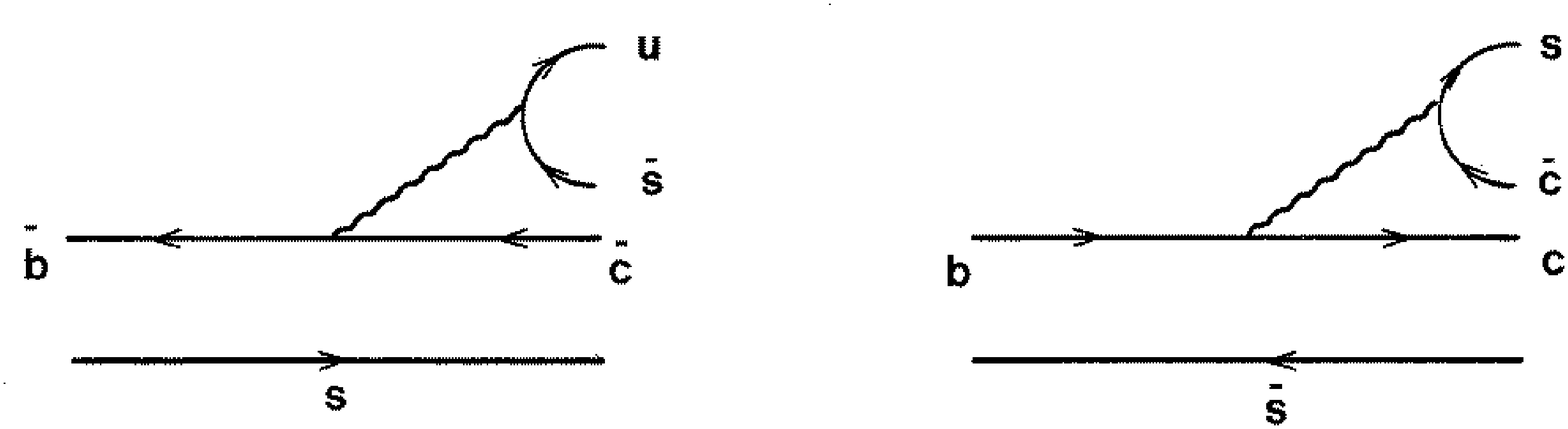}{Tree quark diagrams for $B_s$ and $\bar{B}_s$
leading to $D_s^- \, K^+$.}{figura14}

 Tree quark diagrams for $B_s$ and $\bar{B}_s$
leading to $D_s^- \, K^+$

In order to resolve a twofold ambiguity in the extraction of
$\gamma$ and $\Delta$ from the experiment, an important role of a 
term like sinh ($\Delta \Gamma_s t/2)$ is needed. For the $B_s$-system,
$\Delta \Gamma_s/ \bar{\Gamma}_s \sim 0.1$.

ii) Use of Decay to Self-Tagging final states, which
are flavour-specific \cite{MGW}.

In this case one can use the rate asymmetries as for charged B's \cite{JBU},
with only rates and no time-dependence.

Take the six decays

\begin{center}
\begin{tabular}{ l l l l l l l}
 & & &\ \                \rule{0.25mm}{3mm}
 \raisebox{1.75mm}{\hspace*{-1.75mm}$\longrightarrow$}\ \
 \raisebox{1.75mm}{$K^+$\ $\pi^-$}&&&\\
$B^0$&$\longrightarrow$&$D^0$&$K^{*0}$&&+&c.c.\\
& & \ \                \rule{0.25mm}{3mm}
 \raisebox{-1.25mm}{\hspace*{-1.9mm}$\longrightarrow$}&
 \raisebox{-1.25mm}{$K^-$\ $\pi^+$}&&&\\
 &&&&&&\\
  & & &\ \                \rule{0.25mm}{3mm}
 \raisebox{1.75mm}{\hspace*{-1.75mm}$\longrightarrow$}\ \
 \raisebox{1.75mm}{$K^+$\ $\pi^-$}&&&\\
$B^0$&$\longrightarrow$&$\bar{D}^0$&$K^{*0}$&&+&c.c.\\
& & \ \                \rule{0.25mm}{3mm}
 \raisebox{-1.25mm}{\hspace*{-1.9mm}$\longrightarrow$}&
 \raisebox{-1.25mm}{$K^+$\ $\pi^-$}&&&\\
 &&&&&&\\
   & & &\ \                \rule{0.25mm}{3mm}
 \raisebox{1.75mm}{\hspace*{-1.75mm}$\longrightarrow$}\ \ 
 \raisebox{1.75mm}{$K^+$\ $\pi^-$}&&&\\
$B^0$&$\longrightarrow$&$D_1$&$K^{*0}$&&+&c.c.\\
& & \ \                \rule{0.25mm}{3mm}
 \raisebox{-1.25mm}{\hspace*{-1.9mm}$\longrightarrow$}&
 \raisebox{-1.25mm}{$K^+$\ $K^-$, $\pi^+$\ $\pi^-$}&&&
\end{tabular}
\end{center}

\vspace{0.2cm}

\noindent
where $D_1$ is the $CP = +$ eigenstate for $D^0 - \bar{D}^0$. If the 
corresponding amplitudes  are indicated by $M \, , \,  \bar{M} \,  ,
\, M_+$, the ones for the conjugated processes are $\bar{M} \, , \,
M \, ,  \, M_-$, respectively. The first three amplitudes close
an isospin triangle relation, and thus do the second three
amplitudes. The relative orientation of the two triangles measures
the angle $(2 \gamma)$, where $\gamma$ is that shown in the unitarity
triangle. This contruction is illustrated in Fig. 15,  where
the amplitude $M$ is taken as the reference in the complex plane.

\mxfigura{9cm}{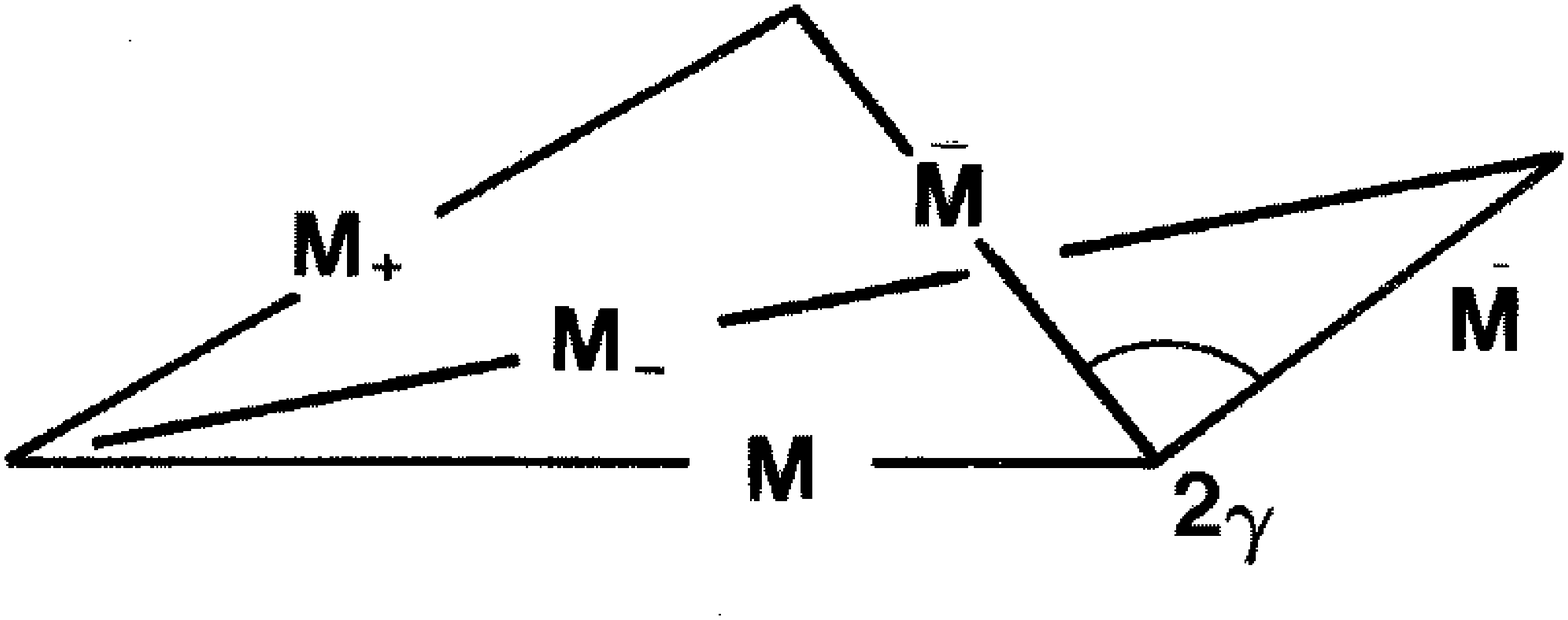}{The two isospin triangle relations 
for the six decays
associated with $B \rightarrow D K^*$.}{figura15}

 The two isospin triangle relations for the six decays
associated with $B \rightarrow D K^*$

\subsection{Outlook}

The flavour structure is one of the main pending questions
in fundamental physics. The Standard Model incorporates the
Cabibbo-Kobayashi-Maskawa mechanism to generate CP-violation. For 3
families, one single CP-phase appears, leading to constrained physics.
A fundamental explanation of the origin of CP-violation is however
lacking.

The CP-odd asymmetries in the $B^0 - \bar{B}_0$ system can lead, by
the choice of appropriate final states, to a determination
of the unitarity triangle when combined with side measurements.
 The three angles $\alpha , \beta$ and $ \gamma = \pi - \alpha - \beta$
can be separately determined and the experimental methods 
 to do so have been discussed
here. But still one perceives that all possible methods for the
determination of the unitarity triangle have not yet
been studied. The hadronic physics involved in some matrix elements needs
further scrutiny. The overall consistency will need of different
experimental methods to extract the sides and angles of the 
unitarity constraint.

\end{document}